\documentclass[usenatbib]{mn2e}
\usepackage{times}
\usepackage{txfonts}
\usepackage{epsfig}
\usepackage{graphicx}
\usepackage{journals}
\usepackage[margin=0.8in, top=1in, bottom=1in, a4paper, pdftex, dvips]{geometry}

\usepackage{graphicx}
\usepackage[usenames,dvipsnames,svgnames,table]{xcolor}
\usepackage{ulem}

\title[Helium in RGB stars]
{Prospects for asteroseismic inference on the envelope helium abundance in red giant stars}

\author[A.-M. Broomhall et al.]{A.-M. Broomhall$^{1,2,3}$\thanks{a-m.broomhall@warwick.ac.uk}, A. Miglio$^{3,4}$, J. Montalb$\rm\acute{a}$n$^5$, P. Eggenberger$^6$,\newauthor W.~J.Chaplin$^{3,4}$, Y. Elsworth$^{3,4}$, R. Scuflaire$^5$, P. Ventura$^7$, G.A. Verner$^3$ \\
$^1$Institute of Advanced Studies, University of Warwick, Coventry, CV4 7HS, UK\\
$^2$Centre for Fusion, Space, and Astrophysics, Department of Physics, University of Warwick, Coventry CV4 7AL, UK\\
$^3$School of Physics and Astronomy,
University of Birmingham, Edgbaston, Birmingham, B15 2TT, UK\\
$^4$Stellar Astrophysics Centre (SAC), Department of Physics and Astronomy, Aarhus University, Ny Munkegade 120, DK-8000, Aarhus C, Denmark \\
$^5$Institut d'Astrophysique et de G$\acute{e}$ophysique de l'Universit$\rm\acute{e}$ de Li$\rm\grave{e}$ge, 17 all$\acute{e}$e du 6 Ao$\rm\hat{u}$t, 4000 Li$\rm\grave{e}$ge, Belgium\\
$^6$Observatoire de Gen$\rm\grave{e}$ve, Universit$\rm\acute{e}$ de Gen$\rm\grave{e}$ve, 51 Ch. des Maillettes, 1290 Sauverny, Switzerland\\
$^7$Osservatorio Astronomico di Roma-INAF, via Frascati 33, I-00040 Monteporzio Catone, Rome, Italy \\}
\date{}

\begin{document}
\maketitle
\begin{abstract} Regions of rapid variation in the internal structure of a star are often referred to as acoustic glitches since they create a characteristic periodic signature in the frequencies of p
modes. Here we examine the localized disturbance arising from the helium second ionization zone in red giant branch and clump stars. More specifically, we determine how accurately and precisely the parameters of the ionization zone can be obtained from the oscillation frequencies of stellar models. We use models produced by three different generation codes that not only cover a wide range of stages of evolution along the red giant phase but also incorporate different initial helium abundances. To study the acoustic glitch caused by the second ionization zone of helium we have determined the second differences in frequencies of modes with the same angular degree, $l$, and then we fit the periodic function described by Houdek \& Gough to the second differences. We discuss the conditions under which such fits robustly and accurately determine the acoustic radius of the second ionization zone of helium. When the frequency of maximum amplitude of the p-mode oscillations was greater than $40\,\rm\mu Hz$ a robust value for the radius of the ionization zone was recovered for the majority of models. The determined radii of the ionization zones as inferred from the mode frequencies were found to be coincident with the local maximum in the first adiabatic exponent described by the models, which is associated with the outer edge of the second ionization zone of helium. Finally, we consider whether this method can be used to distinguish stars with different helium abundances. Although a definite trend in the amplitude of the signal is observed any distinction would be difficult unless the stars come from populations with vastly different helium abundances or the uncertainties associated with the fitted parameters can be reduced. However, application of our methodology could be useful for distinguishing between different populations of red giant stars in globular clusters, where distinct populations with very different helium abundances have been observed. \end{abstract}


\section{Introduction}\label{section[introduction]}

Asteroseismology uses the natural resonant oscillations of stars to study their interiors. With the launch of the \textit{Convection, Rotation and planetary Transits (CoRoT)} and \textit{Kepler} satellites, asteroseismology can now be carried out on a vast and diverse range of stars. Here we study red giant stars which are cool, highly luminous stars that are more evolved than our Sun. However, like our Sun, they have a convective envelope that can stochastically excite acoustic oscillations, which are known as p modes since the main restoring force of these oscillations is a gradient of pressure.

Regions of rapid variation in the internal structure (or sound speed) of a star
create a characteristic periodic signature in the frequencies of p
modes
 \citep[e.g.][]{Vorontsov1988, Gough1990, Basu1994, Roxburgh1994, Basu2004,
Houdek2007}. By analysing this signal, information on the localized
disturbance can be obtained. For example, the periodicity of the signal is related to the sound travel time from that region to the surface \citep[e.g.][]{Vorontsov1988, Gough1990}. Regions of rapid variation, which are known as acoustic glitches, can occur in zones of rapidly changing chemical composition, ionization zones of major chemical elements, and regions where energy transport switches from radiative to convective.

Studies of periodic signatures in the frequencies of solar p modes have allowed the depths of both the convective envelope and the second ionization zone of helium to be determined, the extent of the overshoot at the base of the convection zone to be ascertained and, additionally, the abundance of helium in the envelope to be estimated \citep[e.g.][]{JCD1991, JCD1995, Basu1994, Monteiro1994, Basu1995, Basu1997, Basu1997a, Basu2001, Monteiro2005, JCD2011}. Furthermore, it has been proposed that these techniques can be used to study acoustic glitches in stars other than the Sun \citep[e.g.][]{Monteiro1998, Perez1998, Monteiro2000, Lopes2001, Ballot2004, Basu2004, Verner2004, Verner2006, Houdek2007, Hekker2011} and the first studies of this kind have been conducted \citep[e.g.][]{miglio2010, Mazumdar2012, Mazumdar2014}.

Here we examine the localized disturbance arising from the
second ionization zone of helium, which causes a distinct bump in the
first adiabatic exponent, $\gamma_1$ \citep[e.g.][and references therein]{Basu2004}. The first adiabatic exponent is defined as the logarithm
of the derivative of the pressure ($P$) with respect to the density
$(\rho)$ evaluated at constant entropy $(s)$, i.e.
\begin{equation}\label{equation[gamma 1]}
    \gamma_1=\left(\frac{\textrm{d}\ln
    P}{\textrm{d}\ln\rho}\right)_s.
\end{equation}
$\gamma_1$ can be related to the adiabatic
sound speed, which is assumed to vary only with depth, by
\begin{equation}\label{equation[c gamma]}
    c_\textrm{\scriptsize{s}}^2=\frac{\gamma_1P}{\rho}.
\end{equation}
Therefore, the distinct bump in $\gamma_1$ caused by the
second ionization zone of helium has a corresponding effect on the sound speed
and it is this localized perturbation to $c_\textrm{\scriptsize{s}}$ that causes the
oscillatory signature in the p-mode frequencies.

To investigate the acoustic glitch caused by the second ionization zone of helium we have studied the second differences in the p-mode frequencies, which were defined by \citet{Gough1990} as
\begin{equation}\label{equation[second diffs]}
    \Delta_2\nu_{n,l}\equiv\nu_{n-1, l}-2\nu_{n,l}+\nu_{n+1,l}.
\end{equation}
Here $\nu_{n,l}$ is the frequency of the $n\rm th$ overtone of
the p mode with spherical harmonic degree $l$. Any localized region of rapid variation of the sound speed
will cause an oscillatory component in $\Delta_2\nu_{n,l}$ with a
cyclic frequency of approximately twice the acoustic depth of the
variation. In this paper we characterize the second ionization zone of helium by examining the periodic variations observed in the second differences. The
$\Delta_2\nu_{n,l}$ were used because the first differences are subject to smoothly varying
components which introduce additional parameters to be determined from the fitting process: The mode frequencies are susceptible to near-surface effects that are smoothly varying with frequency, such as the ionization of hydrogen and non-adiabatic processes. These effects are largely reduced by taking the second differences, and this is particularly true close to the frequency at which the amplitude of the oscillations is a maximum ($\nu_{\textrm{\scriptsize{max}}}$), where the trend is approximately flat with frequency. In this paper we have restricted our analysis to modes with frequencies close to $\nu_{\textrm{\scriptsize{max}}}$ and therefore satisfactory results can be obtained simply by fitting a constant offset to the second differences in addition to the oscillatory term caused by the glitch. Furthermore, the second differences are less
susceptible to the propagation of uncertainties in mode frequencies than higher order differences. When determining the second-order differences we require the frequencies of three consecutive overtones. Higher order differences require the frequencies of an increasing number of overtones. In real data the number of overtones for which we can accurately and precisely obtain frequencies will be limited (by the signal-to-noise ratio of the power in the frequency spectrum). Therefore, higher order differences run the risk of there not being enough observable overtones to allow the signature of the second ionization zone of helium to be characterized, and so higher order differences are not suitable for this study.

In stars like the Sun the periodicities caused by the second ionization zone of helium and by abrupt variations in the derivatives of the sound speed
at the base of the convection zone (BCZ) are similar enough that both components must be studied simultaneously. However, in red giant stars, which are the main focus of this study, the base of the convection zone is located deep within the stellar interior (typically at a radius of $0.1R_\star$). Therefore, the acoustic radius of the base of the convection zone and the acoustic radius of the second ionization zone of helium are very different. Furthermore, if the glitch due to BCZ is near the inner turning point of the oscillation, as is the case with red giant branch (RGB) stars, its effect on the oscillation frequencies may not be represented as a periodic component. As a result one can examine the effect of the second ionization zone of helium in isolation, without contamination from the signature of the BCZ.

The main aim of this paper is to determine the effectiveness and robustness of our methodology at characterizing the second ionization zone of helium. We concentrate on two main parameters: the acoustic radius of the second ionization zone of helium and the amplitude of the oscillatory signal at $\nu_\textrm{\scriptsize{max}}$. The location of the ionization zone is important for understanding a star's internal stratification and can potentially be used to help constrain the mass and radius of the star \citep{Mazumdar2005, miglio2010} . The amplitude of the signal at $\nu_\textrm{\scriptsize{max}}$ is a proxy for the helium abundance of the star. With this in mind we have used a wide selection of models to test the limits of our analysis and these models are described in Section \ref{section[models]}. The method by which we examine the periodicity is then described in detail in Section \ref{section[method]}. In Section \ref{section[glitch location]} we discuss the complications that arise when comparing parameters obtained from the p-mode frequencies with those obtained directly from the models. In Section \ref{section[results]} we discuss the reliability of the method at obtaining the acoustic radius of the second ionization zone of helium accurately and robustly. We also discuss the amplitude of the signal at $\nu_\textrm{\scriptsize{max}}$. In Section \ref{section[model comparison]} we compare estimated values of the acoustic depth of the second ionization zone of helium from different models. Finally, concluding summary and discussion is provided in Section \ref{section[discussion]}.

\section{Stellar models}\label{section[models]}

We have examined the robustness of our analysis using the frequencies generated by three different sets of stellar models (and different pulsation codes).
\begin{itemize}
\item{M1.} The first set of models was computed with the \textsc{ATON3.1} code \citep{Ventura2008}, and adiabatic oscillation frequencies of low-degree modes were computed using \textsc{losc} \citep{Scuflaire2008a}. The energy transport in the convective regions was modelled using the classic mixing-length treatment with $\alpha_{\rm MLT} = 1.90$.
We initially assumed the initial heavy-elements mass fraction was $Z=0.020$, and three values for the initial helium mass fraction: $Y=0.250, 0.278$, and 0.400. We therefore explored a rather wide range of $Y$, from slightly larger than current estimates of primordial $Y$ \citep[see][]{Steigman2007, Planck2013} up to values which may be relevant for stars formed in He-enriched environments \citep[see e.g.][and references therein]{Gratton2012}.

In our analysis we considered models of 1.5\,M$_{\odot}$ along the RGB and in the core-helium-burning phase (obtained following the evolution through the helium flash), known as clump stars.
To test the impact of uncertainties in the modelling of near-surface layers on our ability to recover the He-ionization signature, we also considered models computed with $\alpha_{\rm MLT} = 2.05$, adopting the `Full Spectrum of Turbulence'  \citep[FST; ][]{Canuto1996} treatment of convection, and models where the layers above the photosphere were not included in the computation of adiabatic frequencies (see Section \ref{section[glitch location]}). We also considered clump stars generated by this code, with both $Z=0.020$ and $0.001$.

The majority of our analysis is based upon the M1 models, and specifically those with $Y=0.278$, $Z=0.020$, and $\alpha_{\rm MLT} = 1.90$. When the other values of $Y$, $Z$, and $\alpha_{\rm MLT}$ are used it will be stated explicitly.

\item{M2.} We computed the evolution of models of 1.6\,M$_{\odot}$ with the \textsc{geneva}
stellar evolution code \citep{Eggenberger2008}. The
solar mixture of \citet{Grevesse1993} and a solar calibrated value
for the mixing-length parameter were used for these computations. The
initial mass fraction of heavy elements was fixed to $Z=0.02$, while two
different values were used for the initial helium mass fraction: $Y=0.260$
and $0.300$. In order to compare the asteroseismic properties of red
giants with different surface helium abundances, models with similar radii
of 5.6, 6.3, 7.5 and 10.3\,R$_{\odot}$ ascending the RGB were
chosen for the two different values of the helium abundance. The adiabatic
frequencies corresponding to the radial modes of these models were computed
using the \textsc{aarhus} adiabatic pulsation code \citep{JCD2008}.
\item{M3.} A third set of models of 1.6 \,M$_{\odot}$ was computed with the code \textsc{mesa} \citep{Paxton2011}.
We adopted the solar metal mixture by \cite{Grevesse1993}, and a mixing-length parameter of $\alpha_{\rm MLT}=1.90$. We fixed the initial heavy-element mass fraction to $Z=0.020$ and considered  $Y=0.250$ and $0.280$. Adiabatic oscillation frequencies of radial modes were computed using \textsc{losc}.
\end{itemize}

\section{Method}\label{section[method]}

The oscillatory signature of the second ionization zone of helium in the second differences can be described by the following function \citep{Houdek2007}
\begin{equation}\label{equation[fitted function]}
\Delta_2\omega_{n,l}=A\omega_{n,l}\exp(-2b^2\omega_{n,l}^2)\cos\left[2(\tau_{\textrm{\scriptsize{HeII}}}\omega_{n,l}+d)\right]+K,
\end{equation}
where $\omega_{n,l}$ and $\Delta_2\omega_{n,l}$ are the angular
versions of $\nu_{n,l}$ and $\Delta_2\nu_{n,l}$ respectively, $A$ is
the amplitude of the oscillatory component, $\tau_{\textrm{\scriptsize{HeII}}}$ is the acoustic depth of the
second ionization zone of helium, $b$ is the characteristic width of
the region, $d$ is a constant that accounts for the phase of the
signal, and $K$ is a constant offset.

The models used to test our methodology provide frequencies for more
modes than are realistically detectable in asteroseismic
observations (because the signal-to-noise ratio becomes too low).
The underlying power distribution of the observed modes can be
described by a Gaussian centred on the frequency
$\nu_{\textrm{\scriptsize{max}}}$. \citet{Mosser2012} showed that
the full width at half maximum of the Gaussian envelope, $\delta\nu_{\textrm{\scriptsize{env}}}$, for red giant stars is given by
\begin{equation}\label{equation[FWHM gaussian]}
    \delta\nu_{\textrm{\scriptsize{env}}}=0.66\nu_{\textrm{\scriptsize{max}}}^{0.88}.
\end{equation}
Initially, we have only used modes that lie within the range
$\nu_{\textrm{\scriptsize{max}}}\pm0.75\delta\nu_{\textrm{\scriptsize{env}}}$ as we feel this is a reasonable reflection of the number of modes observed in red giant stars \citep[e.g.][]{Carrier2010, DiMauro2011, DiMauro2013, Baudin2012, Deheuvels2012, Jendreieck2012}. However, we do examine the impact on the fit of having both more and less modes available (see Section \ref{section[number modes]}).

We initially assume that frequencies of only radial modes are available. Although frequencies of non-radial modes are observed in real data we will often have to rely on only the radial modes for the following reasons. For the frequencies of the non-radial modes to be perturbed by the ionization zone in a similar manner to that of the $l=0$ modes the observed oscillations must be pure p modes. Frequently, however, the observed non-radial modes are mixed in character, meaning they display properties of both p modes and gravity (g) modes, and their frequencies may significantly deviate from the p-mode asymptotic pattern. It may be possible to infer the pure p-mode frequencies via a detailed modelling of the observed mixed-modes patterns \citep[see e.g.][]{JCD2012}. Whether the inferred acoustic frequencies are accurate enough to determine properties of acoustic glitches needs to be tested, and is beyond the scope of this work.

The extent to which the modes are trapped within the acoustic cavity also varies, and so it is possible that for some stars the non-radial modes may be dominated by p-mode characteristics. This is typically the case for RGB stars with relatively high luminosities ($L\gtrsim 100$ L$_{\odot}$), in which non-radial modes become effectively trapped in the acoustic cavity, and frequencies are largely insensitive to the structure of the deep interior (see e.g. \citealt{Dupret2009, Montalban2010, Dziembowski2012}). Therefore, we have also considered cases where the frequencies of $l=1$ and $l=2$ modes are expected to be predominantly acoustic in character (see Section \ref{section[results diff l]}).

Since the models do not supply uncertainties with the frequencies, artificial uncertainties were added. We initially gave each second difference an uncertainty of $0.044\,\rm\mu Hz$. This was based on the uncertainties of the central $l=0$ frequencies observed by \citet{DiMauro2011} and \citet{Deheuvels2012}, both of whom used \textit{Kepler} data to obtain frequencies of red giant stars. \citet{DiMauro2011} and \citet{Deheuvels2012} use data sets which were 30 and 365\,d, respectively. However, there are now 4yr of \textit{Kepler} data available and so to make the uncertainties used here appropriate for 1460\,d we scaled by the square root of the ratios of the length of time series \citep{Libbrecht1992}. Using this method the uncertainties estimated by \citet{DiMauro2011} and \citet{Deheuvels2012} produced consistent results and so an average was taken. However, the size of the uncertainties associated with the observed frequencies can vary depending on, for example, the quality of the data, the length of observations, and the intrinsic properties of the modes. Therefore we have also tested how the results are affected if the uncertainties were both larger and smaller than this (see Section \ref{section[results diff errors]}).

\begin{figure}
    \centering
  \includegraphics[width=0.4\textwidth, clip]{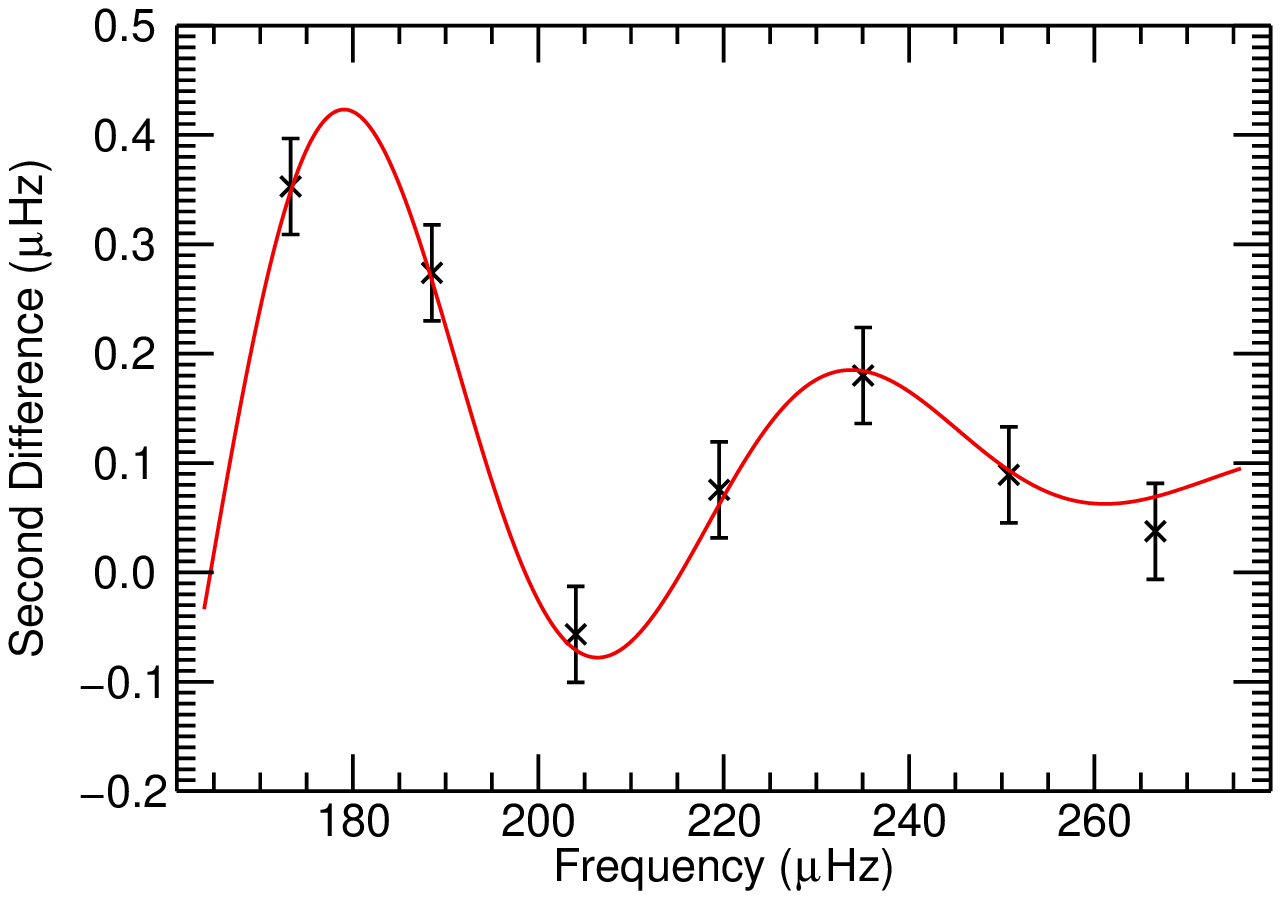}\\
  \includegraphics[width=0.4\textwidth, clip]{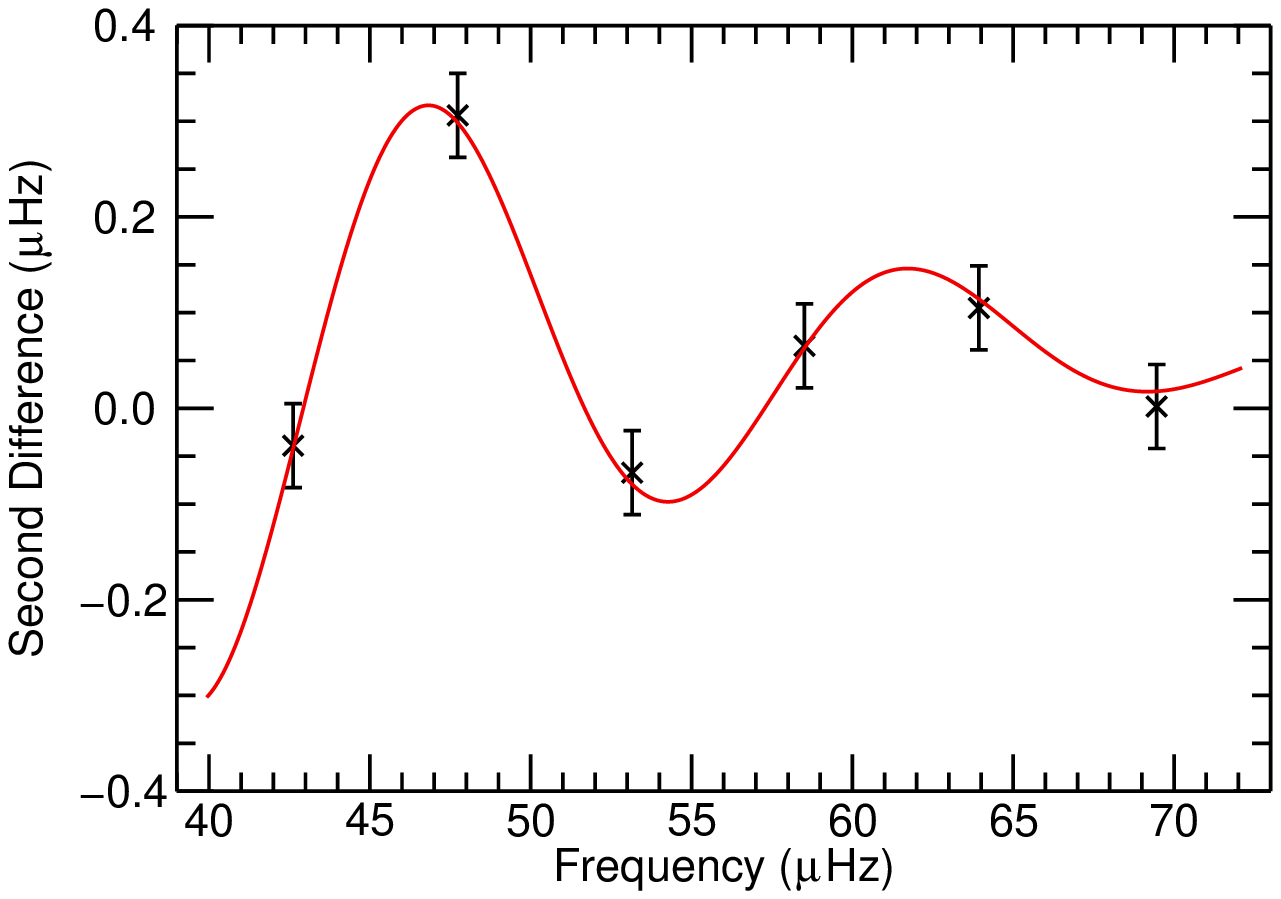}\\
  \caption{Examples of the $l=0$ second differences (black crosses) predicted by two different M1 RGB models as a function of frequency. The fitted function is represented by the red line. In the top panel $\nu_{\textrm{\scriptsize{max}}}=215.60\,\rm\mu Hz$, while in the bottom panel $\nu_{\textrm{\scriptsize{max}}}=54.72\,\rm\mu Hz$.}\label{figure[eg fit]}
\end{figure}

An example of the model second differences predicted for two different stars and the resultant fitted function can be seen in Fig. \ref{figure[eg fit]}. The two models have well-separated $\nu_{\textrm{\scriptsize{max}}}$ meaning they represent different stages of evolution along the RGB.

\subsection{Errors associated with a non-linear least-squares fit of correlated data}\label{section[errors]}
We have fitted equation (\ref{equation[fitted function]}) to the data using a non-linear least squares fit. However, this method does not
take account of the fact that the data points will be correlated.
Although this will not alter the values of the fitted
parameters it can have an effect on the formal errors
associated with the fitted parameters.

The analysis performed in this paper relies on the determination of three key parameters: the acoustic depth of the second ionization zone of helium, $\tau_{\textrm{\scriptsize{HeII}}}$, the amplitude of the oscillatory signal, $A$, and the acoustic radius of the star, $T$. Although, $\tau_{\textrm{\scriptsize{HeII}}}$ and $A$ are determined by fitting equation (\ref{equation[fitted function]}) to the frequency differences, $T$ must be obtained separately. The acoustic radius of the star, $T$, is given by
\begin{equation}\label{equation[acoustic radius]}
    T=\int^R_0\frac{\textrm{d}r}{c_\textrm{\scriptsize{s}}}=\frac{1}{2\Delta\nu},
\end{equation}
where $R$ is the radius of the star, $c_\textrm{\scriptsize{s}}$ is the sound speed, and $\Delta\nu$ is the large separation of the frequencies, or the difference in frequency of consecutive overtones of modes with the same harmonic degree. The large separation can be obtained by determining the gradient of the linear fit between radial order, $n$, and mode frequency \citep[see][and references therein for more details]{Broomhall2011}.

To correct the uncertainties on the fitted parameters for correlations in the second differences we ran 20 000 Monte Carlo simulations where the
frequencies were varied by a randomly determined amount
$\delta\nu_{n,l}$ from their model values such that
\begin{equation}\label{equation[vary freqw]}
\nu_{n,l}'=\nu_{n,l}+\delta\nu_{n,l},
\end{equation}
where $\delta\nu_{n,l}$ was determined using a random number
generator that supplied a number from a normal distribution with a
mean value of zero and a standard deviation that was determined by
the error on the frequency. For each Monte Carlo simulation, equation
(\ref{equation[fitted function]}) was fitted to the adjusted
frequencies, $\nu_{n,l}'$.

Using Monte Carlo simulations to determine the errors on the fitted parameters has the added advantage of providing a measure of the robustness of the original fit, since the distribution of the obtained parameters should, ideally, be Gaussian. For each model, we defined the robustness as one of three options: robust, mostly robust, or not robust. The classification depended on how similar the distributions of the parameters obtained from the Monte Carlo simulations were to a Gaussian. In order to classify the robustness of the fit we considered three tests, which we will now describe.
\begin{enumerate}
  \item Let $P_{\textrm{\scriptsize{orig}}}$ be the value found for one of the parameters in the original fit, and $\sigma_{\textrm{\scriptsize{orig}}}$ be the uncertainty associated with that parameter. Then let $P_i$ be the value found for that same parameter in one of the Monte Carlo simulations, and $\sigma_i$ be its associated uncertainty. For each simulation, $1\le i\le20,000$, we determine $(P_i-P_{\textrm{\scriptsize{orig}}})/(\sigma_i^2+\sigma_{\textrm{\scriptsize{orig}}}^2)^{1/2}$. We then determine the mean, standard deviation, and a histogram of this value. If the distribution is a Gaussian one would expect 68.3 percent of the simulations to lie within 1 standard deviation of the mean. To allow for statistical deviations we require between 70.6 and 65.8 percent of the simulations to lie within 1 standard deviation of the mean (this corresponds to between 1.05 and 0.95 standard deviations in a true Gaussian distribution).
  \item We require that over 70 per cent of the simulations were successfully fit.
  \item We fit a Gaussian to the observed histogram and ensured there were no large differences, such as the presence of a bimodal distribution or significant skew in the observed distribution.
\end{enumerate}
For the fit to be classified as robust all three parameters were required to pass all three tests. If, for any of the three parameters, the test (i) failed but tests (ii) and (iii) were passed we classified the result as mostly robust. This means that although the fit is not totally robust we still trust the uncertainties implied by the simulations. Finally, if test (i) failed and either/both of tests (ii) and (iii) failed we classified the fit as not robust. We note that, when considered on its own the acoustic radius of the star was always considered to be robust. Therefore, if any of the simulations were classified as mostly robust or not robust it was because either $\tau_\textrm{\scriptsize{HeII}}$ or $A$ could not be obtained robustly.

If the fit was deemed to be robust or mostly robust the uncertainties associated with the original fit were scaled by multiplying by the ratio of the standard deviation on $P_i$ and the mean of $\sigma_i$.

\section{Location of the glitch: acoustic radius and depth}\label{section[glitch location]}
The location of the glitch provides information on the chemical composition and stratification of the stellar interior. However, the expression used to infer the location of the glitch due to helium ionization (equation \ref{equation[fitted function]}) is based on several simplifying assumptions. A critical appraisal of its limitations is therefore needed when comparing the location of the glitch in models to that inferred from the frequencies.

Here, we discuss briefly a few issues that we consider relevant.
\begin{itemize}
\item Is the asymptotic expression of the eigenfunctions (used to derive equation \ref{equation[fitted function]}) accurate enough for the relatively low-radial-order acoustic modes observed in giant stars?
\item How is the acoustic surface of the star/model defined, i.e. where is the origin of the coordinate $\tau$\,?
\item How is our inference on the location of the glitch affected by the cut-off frequency, and by uncertainties on the modelling of near-surface layers?
\item How is the definition of the acoustic glitch related to helium ionization?
\end{itemize}

Before discussing some of these issues in the case of models of giant stars, we shall however recall  \citep[see also][]{Houdek2007}  that the precision of the asteroseismic calibrations does not rely directly on the precision of the asymptotic expression. The latter is used to design indicators that we believe are less biased by `known unknowns' (e.g. modelling of near-surface layers), but comparisons between observed and model frequencies need not be undermined by shortcomings of approximated expressions. 
Moreover, the primary aim of our work is to evaluate under which circumstances we expect to detect the signature of helium ionization in giants. 

\begin{figure}
\centering
  \includegraphics[height=0.45\textwidth, clip, angle=-90]{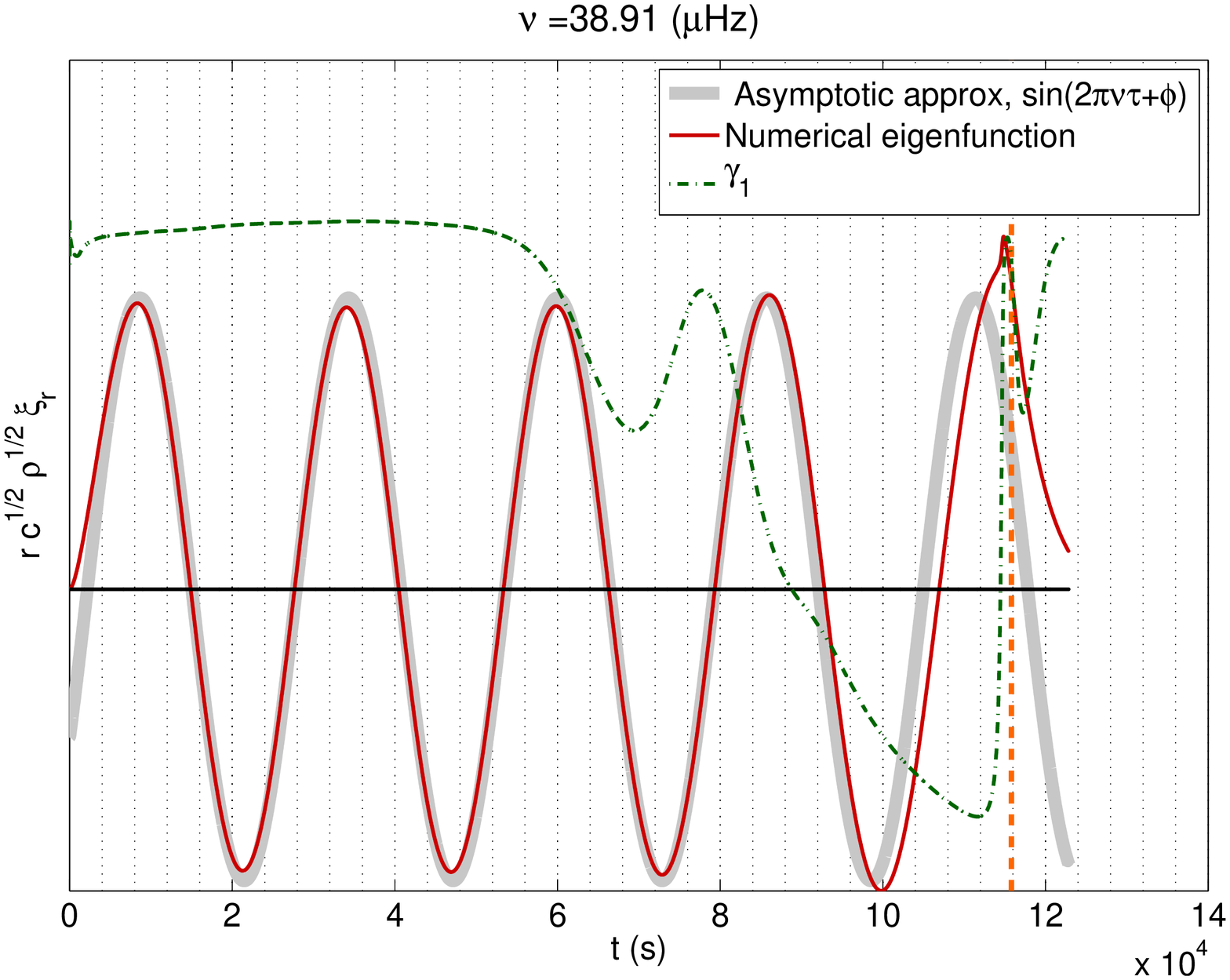}
  \includegraphics[height=0.45\textwidth, clip, angle=-90]{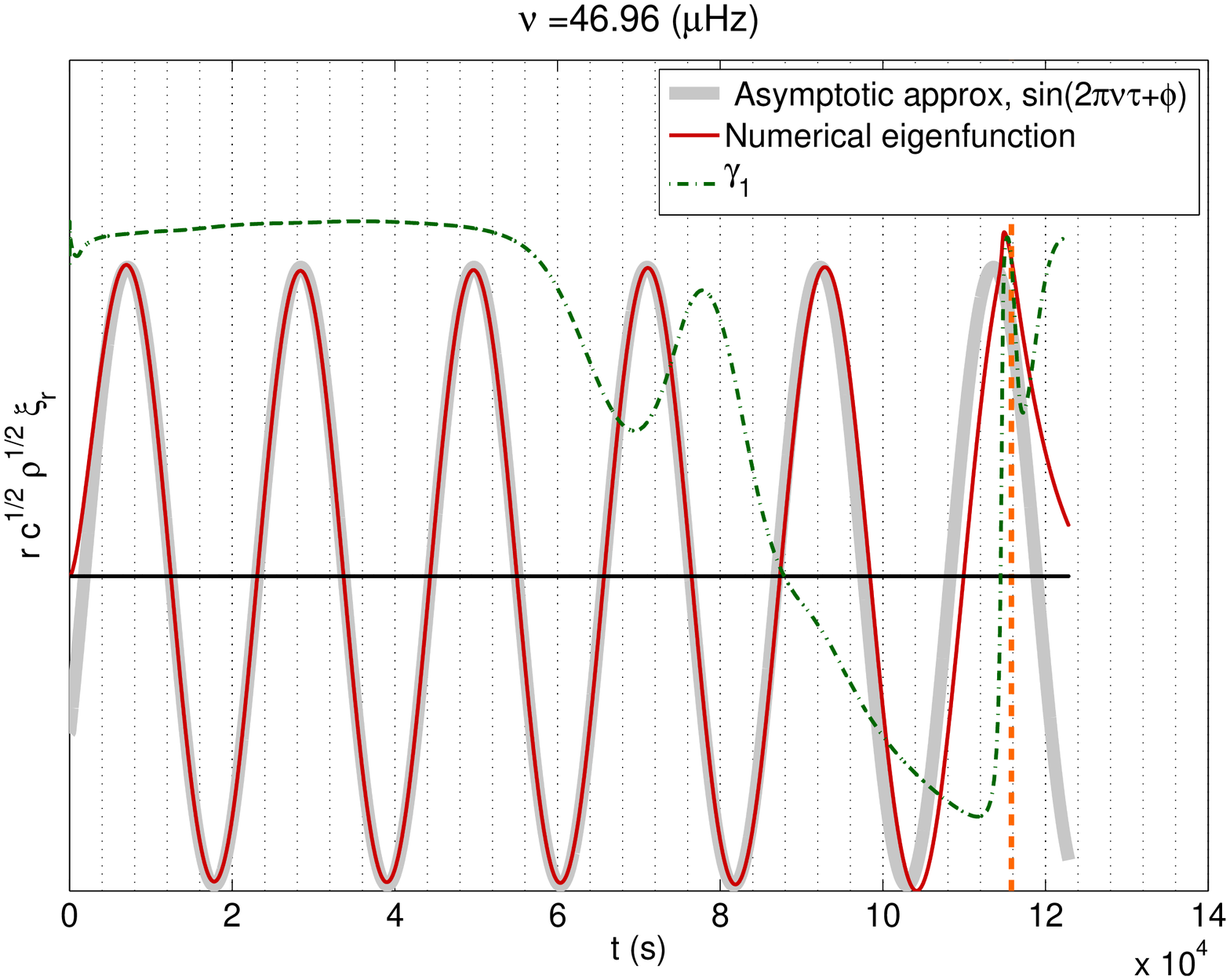}
  \caption{Red line: behaviour of the scaled radial displacement eigenfunction of two radial modes (top panel: $\nu=38.91$ $\mu$Hz; bottom panel: $\nu=46.69$ $\mu$Hz) in a 1.5 M$_\odot$, $R=11.9$ R$_{\odot}$ model. Thick grey lines: asymptotic approximation of the eigenfunctions. Dashed green line: first adiabatic exponent. Orange dashed line: Demonstrates the location of the photosphere.}\label{figure[eigen]}
\end{figure}

In giants the second ionization zone of helium is located far from turning points and well within the almost adiabatically stratified interior. In that region the vertical wavenumber may be well approximated by neglecting the acoustic cutoff frequency ($\omega_{\rm a}$) and the asymptotic phase function is a good approximation of the numerical solution (see Fig. \ref{figure[eigen]}).
Furthermore, in that region the shift in the phase of the eigenfunctions due to $\omega_{\rm a}$ is found to be negligible and the simple asymptotic approximation of the radial eigenfunctions (see equation 9 in \citealt{Houdek2007}) gives a remarkably accurate representation of the numerically computed radial displacement eigenfunction. In the model shown in Fig. \ref{figure[eigen]}, in the helium second ionization region a phase shift of $\lesssim 1\%$ is found when considering modes of different frequency ($\nu \simeq 38$ and $\simeq 47$ $\mu$Hz).

The inferred acoustic depth of the glitch will nonetheless be affected by the behaviour of modes in (and the properties of) near-surface layers \citep[see][]{Monteiro1994, JCD1995}. This, however, can be partly taken into account following the approach suggested by \citet{Ballot2004} and \citet{Mazumdar2005}. The authors suggest to use a quasi-unbiased indicator of the acoustic radius of the sharp feature, obtained by subtracting the acoustic depth from the total acoustic radius determined using the average large frequency separation. We refer to this quantity as the acoustic radius of the glitch, $t_{\textrm{\scriptsize{HeII}}}$.

This indicator is almost unbiased by both surface effects and by the definition of the acoustic surface (see below), and therefore in what follows we compare the acoustic radius of the glitch as inferred from the model frequencies with that determined directly from the model structure \citep[as in][]{Ballot2004}. 

Comparisons of the acoustic depths of glitches inferred from the frequencies with those of stellar models depend on the definition of the acoustic surface of the star, i.e. the origin of the variable $\tau$.
Some authors (e.g. \citealt{Monteiro2005}) adopt the photospheric radius as the origin of the acoustic depth, while \citet{Houdek2007} advocate for a much more extended surface, which, for example, in the case of the Sun extends $\sim 200$ s above the photosphere \citep{Houdek2007}. This clearly has a significant impact ($\sim$ 5\%) on the estimate of the depth of the HeII ionization region.
Moreover, since $\tau$ depends crucially on the near-surface properties of the star, any comparisons between acoustic depths in stellar models and those estimated from observed frequencies will be significantly affected by our shortcomings in modelling near-surface layers.

Finally, the definition of the acoustic glitch itself deserves a further look.
We have considered, as in \citet{Houdek2007}, the local variation of $\gamma_1$ as the main origin of the acoustic glitch associated with helium ionization. Since \citet{Houdek2007} were looking for a description of the glitch which is directly related to the helium abundance, they considered, as a smooth model, a model with no helium, leading to a definition of the glitch in the second ionization zone of helium as a local depression in $\gamma_1$.

\begin{figure}
\centering
  \includegraphics[width=0.4\textwidth, clip]{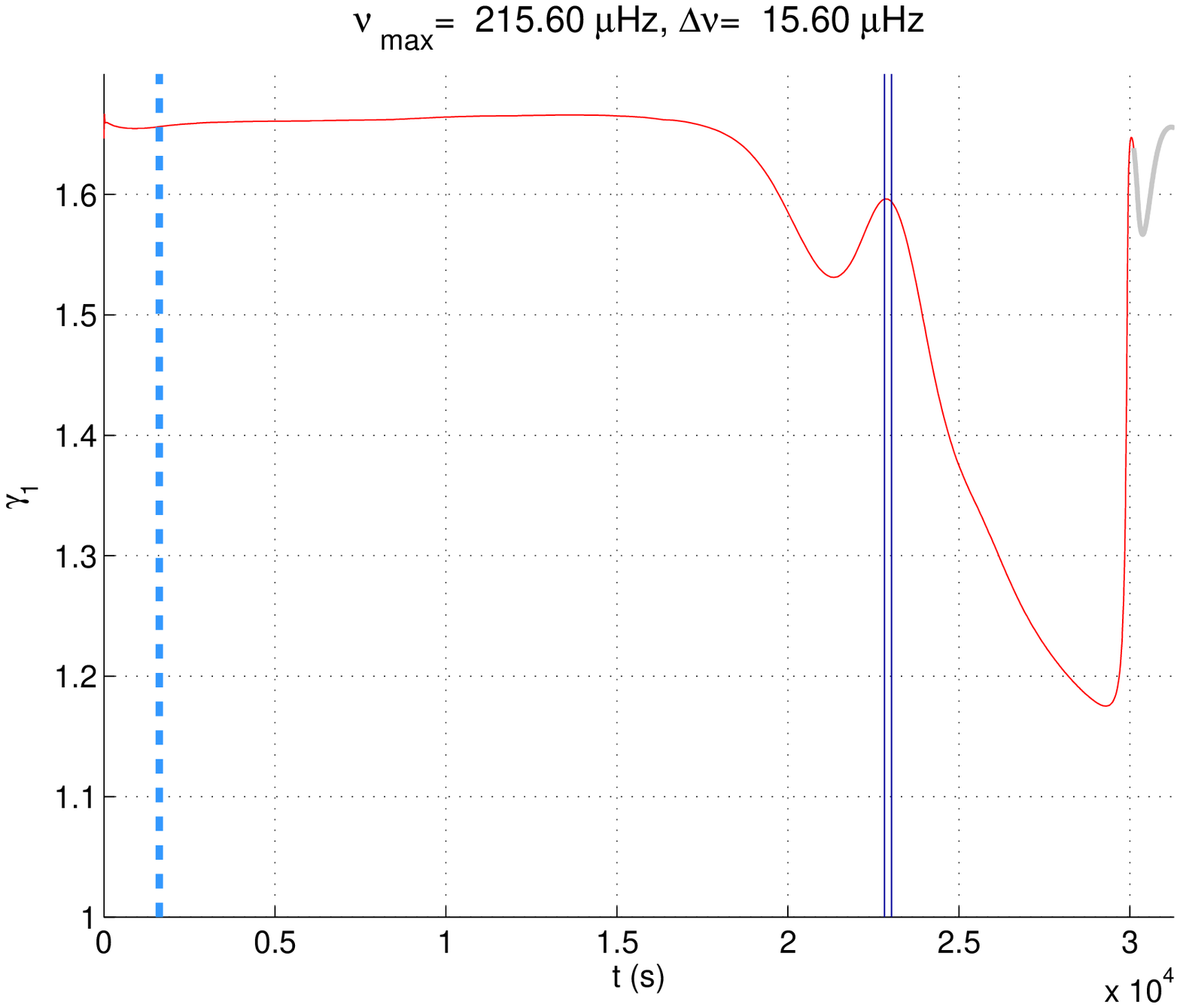}\\
  \includegraphics[width=0.4\textwidth, clip]{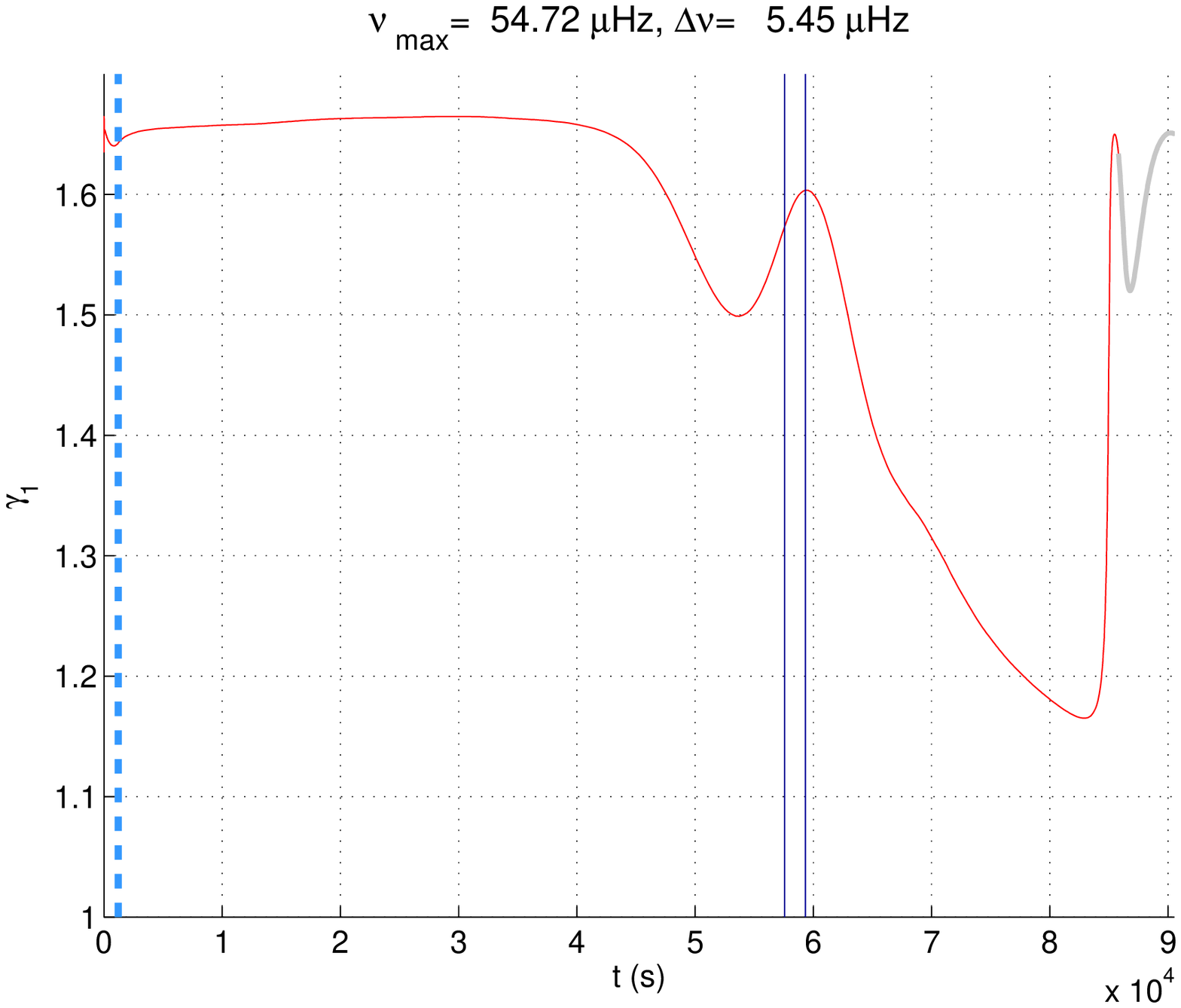}\\
  \caption{ Variation in the first adiabatic exponent as a function of acoustic radius (the layers in the atmosphere are shown in grey).  Vertical solid lines indicate the $1\sigma$ region of the acoustic radius of the glitch obtained from fitting the simulated data.
The latter is in good agreement with the local maximum of $\gamma_1$ in between the regions of first and second ionization of helium.
   The two panels shown here correspond to the two examples plotted in Fig. \ref{figure[eg fit]}.
Vertical dashed lines indicate the acoustic radius of the base of the convective envelope.}\label{figure[eg model]}
\end{figure}

Since the glitch is a rather extended feature in $t$, we note that there is a possible ambiguity concerning the definition of the smooth fictitious acoustic structure: the latter can also be obtained by removing the local maximum in $\gamma_1$ in between the first and second helium ionization zones.
Further tests are needed, however, to ascertain which fictitious smooth model (hence definition of the acoustic glitch) is more appropriate to the oscillation modes relevant to this study, and whether other approximations leading to expression (\ref{equation[fitted function]}) may contribute to inaccuracies in such an expression. In our tests, however, we find that the acoustic radius of the glitch  as determined by fitting the frequencies appears to be consistent with that of a local maximum of $\gamma_1$ in the models, as illustrated in Fig. \ref{figure[eg model]}. This discrepancy could be due to the simplifying assumptions made in associating equation (\ref{equation[fitted function]}) with the true form of the glitch in $\gamma_1$. For example, we have not taken into consideration the effect of the first ionization zone of helium. We remind the reader once again that although the asymptotic expression may be inaccurate, the precision of asteroseismic calibrations does not rely directly on it.

Given the above discussion, in future sections when we compare the acoustic radius of the ionization zone determined by fitting the frequencies with that determined directly from models, the model value is defined as the local maximum of $\gamma_1$.

To quantify the impact of some of the above-mentioned uncertainties and approximations, we have conducted tests on how sensitive the inferred properties of the glitch are to uncertainties in the properties of near-surface layers.
First, we considered a model of $M=1.5 M_\odot$, with  $R=11.9$ R$_\odot$. We then constructed a second model by removing the layers above the photosphere from the first model. We computed frequencies for both models and fitted the periodic component in the second frequency differences. Although in the two cases considered we obtain a different location of the glitch in terms of its acoustic depth, we find the same (within uncertainties) acoustic radius.

\begin{figure}
\centering
  \includegraphics[width=0.45\textwidth, clip]{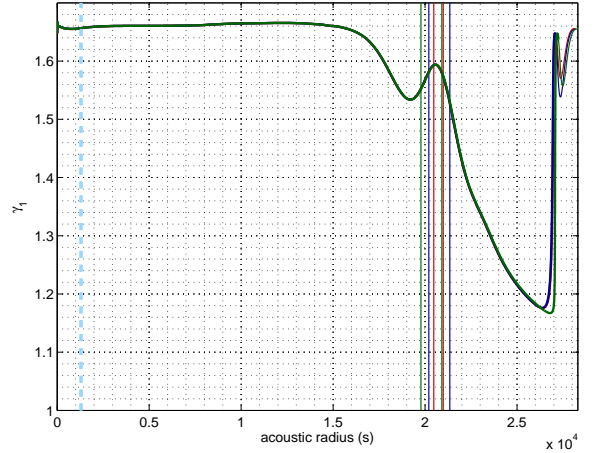}
  \caption{As in Fig. \ref{figure[eg model]}, but for  $M=1.375$ M$_\odot$, $R=3.04$ R$_\odot$ models computed with different assumptions about the efficiency of energy transport by convection: green: FST; blue: MLT with $\alpha_{\rm MLT}=1.90$; red: MLT with $\alpha_{\rm MLT}=2.05$.}\label{figure[convection]}
\end{figure}

Similar results were obtained considering three \textsc{aton} (M1) models of $M=1.375$ M$_\odot$, $Z=0.027$, $Y=0.280$, a photospheric radius $R=3.04$ R$_\odot$, and computed with different prescriptions for the energy transport in convective regions: $\alpha_{\rm MLT}=1.90, 2.05$\footnote{Since these two models have the same radius, their $T_{\rm eff}$ differs.}, and with FST \citep{Canuto1996}, adopting a \textit{fine-tuning parameter} $\alpha^*= 0.16$. The acoustic radius of the glitch in $\gamma_1$ in these models (see Fig. \ref{figure[convection]}) is the same. No significant difference (given the quoted uncertainties) is found when inferring the location of the glitch from the frequencies. No significant difference is found either in the estimated amplitude of the component.

%

As suggested above, and checked in tests presented below, comparing the location of the glitch in terms of the estimated acoustic radius rather than depth provides an indicator which is less biased by the assumed definition of the acoustic surface of the star.

\section{What can be reliably obtained from the fitted results?}\label{section[results]}

\begin{figure}
\centering
  \includegraphics[width=0.4\textwidth, clip]{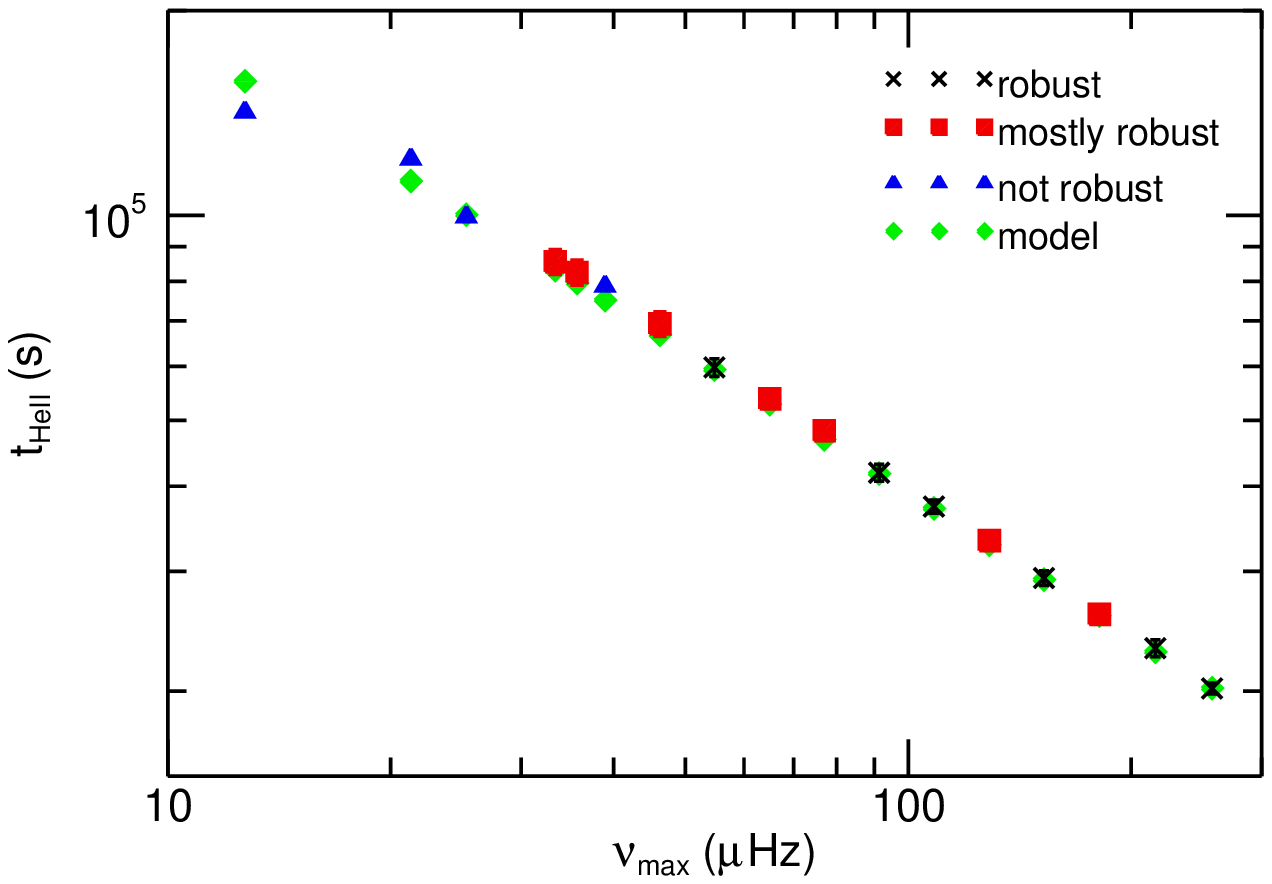}\\
  \includegraphics[width=0.422\textwidth, clip]{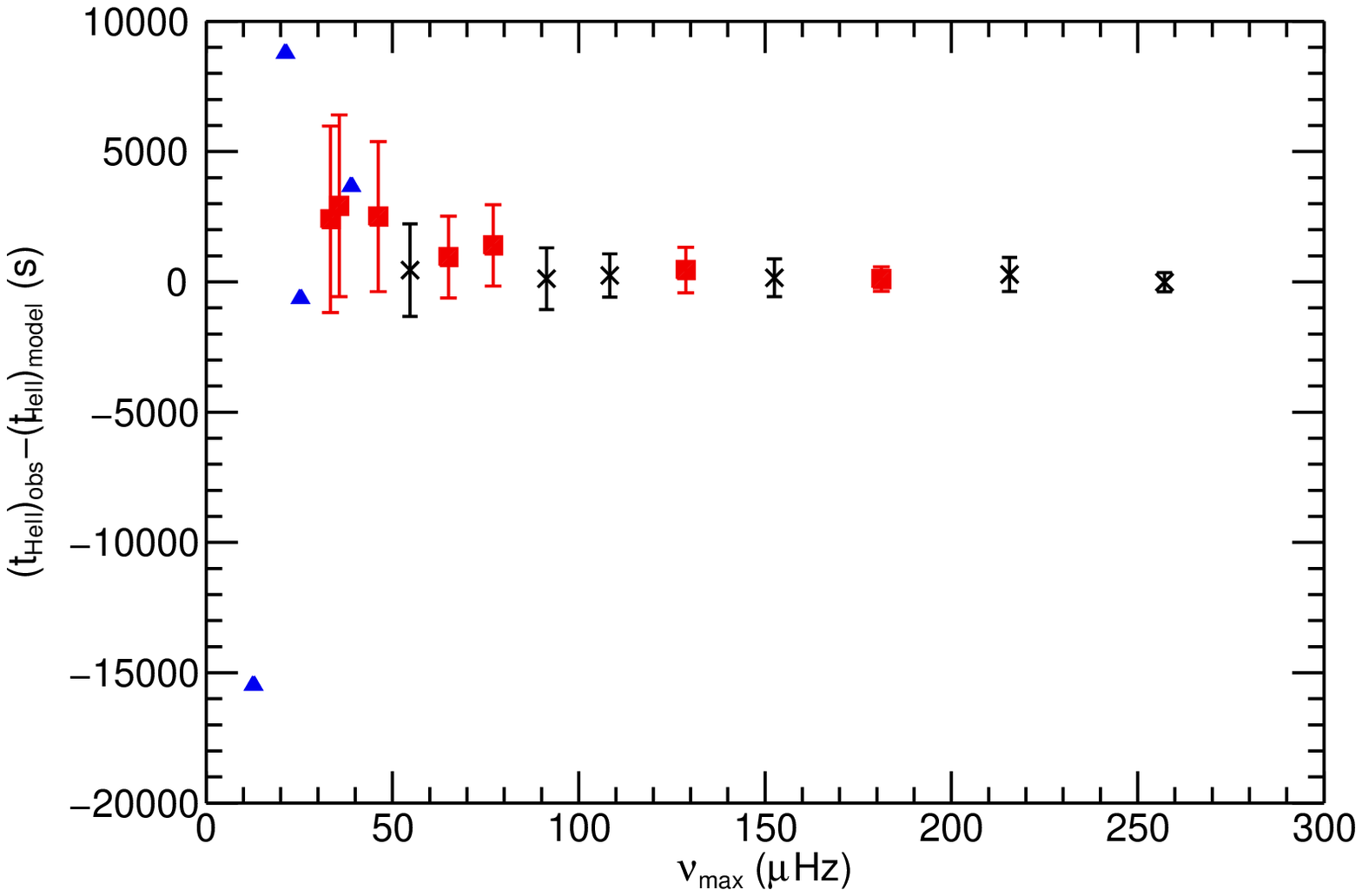}\\
  \caption{Top panel: acoustic radius of the second ionization zone of helium, $t_{\textrm{\scriptsize{HeII}}}$, as a function of $\nu_{\textrm{\scriptsize{max}}}$. The errorbars associated with the robust and mostly robust fits are typically of the same magnitude as or smaller than the plotted symbols. No errorbars have been included for the not robust fits as we do not consider these uncertainties to be well defined. The M1 models with $Y=0.278$ and $Z=0.020$ were used and all of the modelled stars belong to the RGB. Bottom panel: difference between the fitted and model $t_{\textrm{\scriptsize{HeII}}}$ as a function of $\nu_{\textrm{\scriptsize{max}}}$. The symbols and colours represent the same as in the top panel of this figure. }\label{figure[numax vs depth]}
\end{figure}

Fig. \ref{figure[numax vs depth]} shows results from fits to our model data. Plotted in the top panel is the best-fitting acoustic radius of the second ionization zone of helium, $t_{\textrm{\scriptsize{HeII}}}$, as a function of $\nu_{\textrm{\scriptsize{max}}}$.
At higher values of $\nu_{\textrm{\scriptsize{max}}}$ (i.e. above $\sim40\,\rm\mu Hz$) the fit is either robust or mostly robust and follows a smooth trend. When $\nu_{\textrm{\scriptsize{max}}}<40\,\rm\mu Hz$ the robustness of the fits deteriorates. The bottom panel of Fig. \ref{figure[numax vs depth]} indicates that the $t_{\textrm{\scriptsize{HeII}}}$ obtained for all of the fits that are either robust or mostly robust lie within $1\sigma$ of the model $t_{\textrm{\scriptsize{HeII}}}$. These results, therefore, indicate the strength of this method for accurately obtaining $t_{\textrm{\scriptsize{HeII}}}$.

\begin{figure}
\centering
  \includegraphics[width=0.4\textwidth, clip]{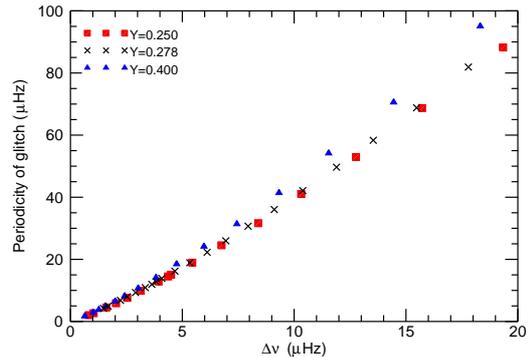}\\
  \caption{Periodicity of the acoustic glitch caused by the second ionization zone of helium as a function of $\Delta\nu$, for various M1 RGB star models with different initial helium abundances (see legend).}\label{figure[model radius vs deltanu]}
\end{figure}

As we have shown it is only at low $\nu_{\textrm{\scriptsize{max}}}$ where the method becomes less reliable. There are three main reasons for this.
\begin{enumerate}
  \item The number of overtones used to perform the fit decreases with $\nu_{\textrm{\scriptsize{max}}}$. We recall that the number of overtones used to fit the acoustic glitch was defined by the width of the Gaussian envelope, using equation (\ref{equation[FWHM gaussian]}). When $\nu_{\textrm{\scriptsize{max}}}<100\,\rm\mu Hz$ only six overtones are used. This is only one more than the number of unknowns we are trying to fit (see equation \ref{equation[fitted function]}).
  \item The maximum amplitude of the envelope of the oscillatory signal decreases with $\nu_\textrm{\scriptsize{max}}$ therein making the signal harder to fit at low $\nu_\textrm{\scriptsize{max}}$.
  \item The function becomes difficult to fit at low $\nu_{\textrm{\scriptsize{max}}}$ because the periodicity caused by the second ionization zone of helium is similar in magnitude to the large separation of the frequencies, $\Delta\nu$, i.e. the resolution of the second differences. Fig. \ref{figure[model radius vs deltanu]} shows the variation of $t_{\textrm{\scriptsize{HeII}}}$ as a function of the large separation, $\Delta\nu$, which also describes the resolution of the data points we are trying to fit when only the radial modes are used. At high $\nu_{\textrm{\scriptsize{max}}}$, the periodicity of the acoustic glitch is large enough compared with %
      $\Delta\nu$ that equation (\ref{equation[fitted function]}) can be fitted robustly. However, as we move to lower $\nu_{\textrm{\scriptsize{max}}}$ the gradient of the curve becomes shallower. Therefore the periodicity of the glitch is no longer large compared with $\Delta\nu$ and, consequently, equation (\ref{equation[fitted function]}) is harder to fit. Including non-radial modes in the analysis would reduce the resolution of the second differences and so, in theory, make the signal easier to fit. However, as we discuss in the next section, including non-radial modes does not always result in more accurate and robust results. We note that $\nu_{\textrm{\scriptsize{max}}}<40\,\rm\mu Hz$ corresponds to $\Delta\nu<4.2\,\rm\mu Hz$.
\end{enumerate}

\begin{figure}
    \centering
  \includegraphics[width=0.4\textwidth, clip]{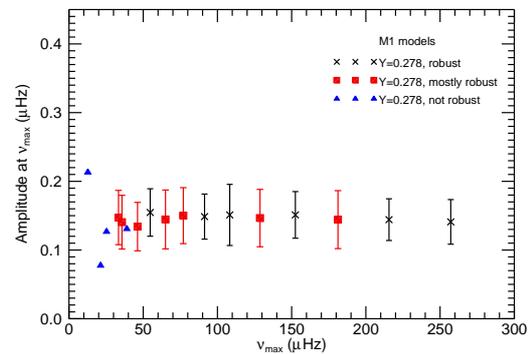}\\
  \caption{Amplitude of the signal from the helium second ionization code as a function of $\nu_{\textrm{\scriptsize{max}}}$. Here the results for the M1 models for RGB stars with $Y=0.278$ and $Z=0.020$ are plotted.}\label{figure[amplitudes]}
\end{figure}

In addition to the acoustic depth of the second ionization zone of helium, the amplitude of the envelope of the signal at $\nu_{\textrm{\scriptsize{max}}}$ can also be extracted from the fitted parameters. The amplitude of the signal at $\nu_{\textrm{\scriptsize{max}}}$ is given by
\begin{equation}\label{equation[amp at numax]}
    A_{\textrm{\scriptsize{max}}}=A\nu_{max}\exp(-2b^2\nu^2_{n,l}),
\end{equation}
and is expected to be correlated with
 the helium abundance in the
helium ionization zone. $A_{\textrm{\scriptsize{max}}}$ is plotted as a function of frequency in Fig. \ref{figure[amplitudes]} and is observed to be relatively stable with $\nu_{\textrm{\scriptsize{max}}}$, particularly when $\nu_{\textrm{\scriptsize{max}}}\gtrsim50\,\rm\mu Hz$. This indicates that the amplitude of the oscillatory signal is larger with respect to the mode frequencies when $\nu_{\textrm{\scriptsize{max}}}$ is smaller.

\subsection{The effect on the fitted parameters of using $l=1$ and $2$ modes}\label{section[results diff l]}

Since, in real data, it is likely that we will be able to observe $l=1$ and $2$ modes in addition to the radial modes we have tested the effect of using higher degree modes when fitting the data. We note here that in each large separation interval, the spectrum of non-radial oscillations is populated by several modes, many of which have large inertias and are gravity-dominated modes. Here we only consider those non-radial modes with the lowest inertias in each $\Delta\nu$ range as these are the modes that are most similar to p modes. However, care must be taken because, as we will see later in this section, the frequencies of some of these modes are significantly perturbed by the interaction with gravity modes. To determine whether using the $l=1$ and $2$ modes `improved' the fitted results we must first define what we mean by an improvement. We considered two aspects.
\begin{enumerate}
  \item The robustness of the fit: it is important that any obtained values are robust with reliable uncertainties and so if the robustness of the fit was upgraded the fit was considered to have been improved regardless of whether the obtained values were more accurate. However, if the robustness was degraded the fit was not considered to have been improved.
  \item The difference between the model and fitted value of $t_{\textrm{\scriptsize{HeII}}}$: we determined the difference between the fitted and model values of $t_{\textrm{\scriptsize{HeII}}}$ relative to the size of the uncertainties associated with the fitted $t_{\textrm{\scriptsize{HeII}}}$. This indicator was only considered if the robustness classification remained the same and was considered to be `robust' or `mostly robust'. Since the formal errors associated with `not robust' fits are unreliable the relative difference between the observed and model $t_{\textrm{\scriptsize{HeII}}}$ would not be a true indicator of the quality of the fitted result.
\end{enumerate}

\begin{figure}
\centering
  \includegraphics[width=0.4\textwidth, clip]{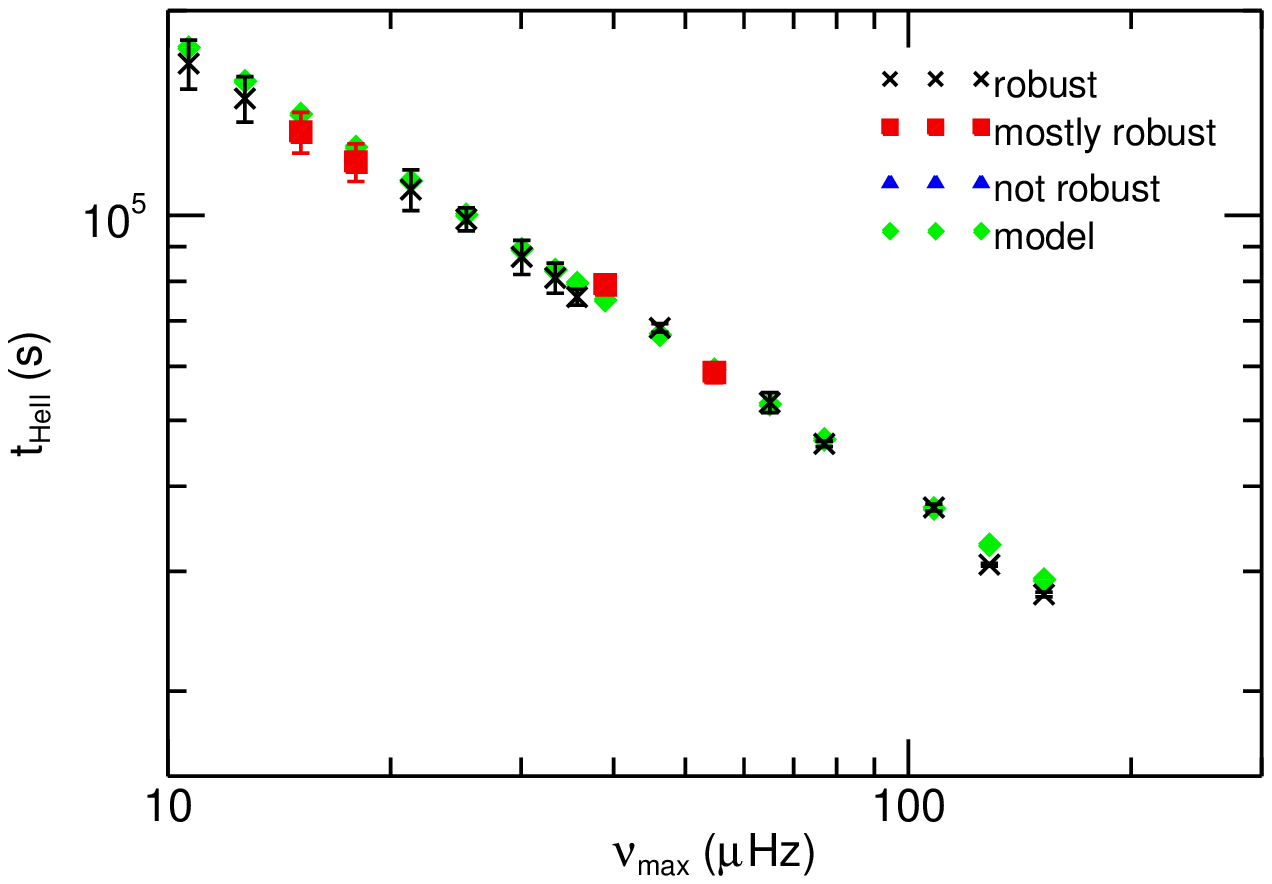}\\
  \includegraphics[width=0.4\textwidth, clip]{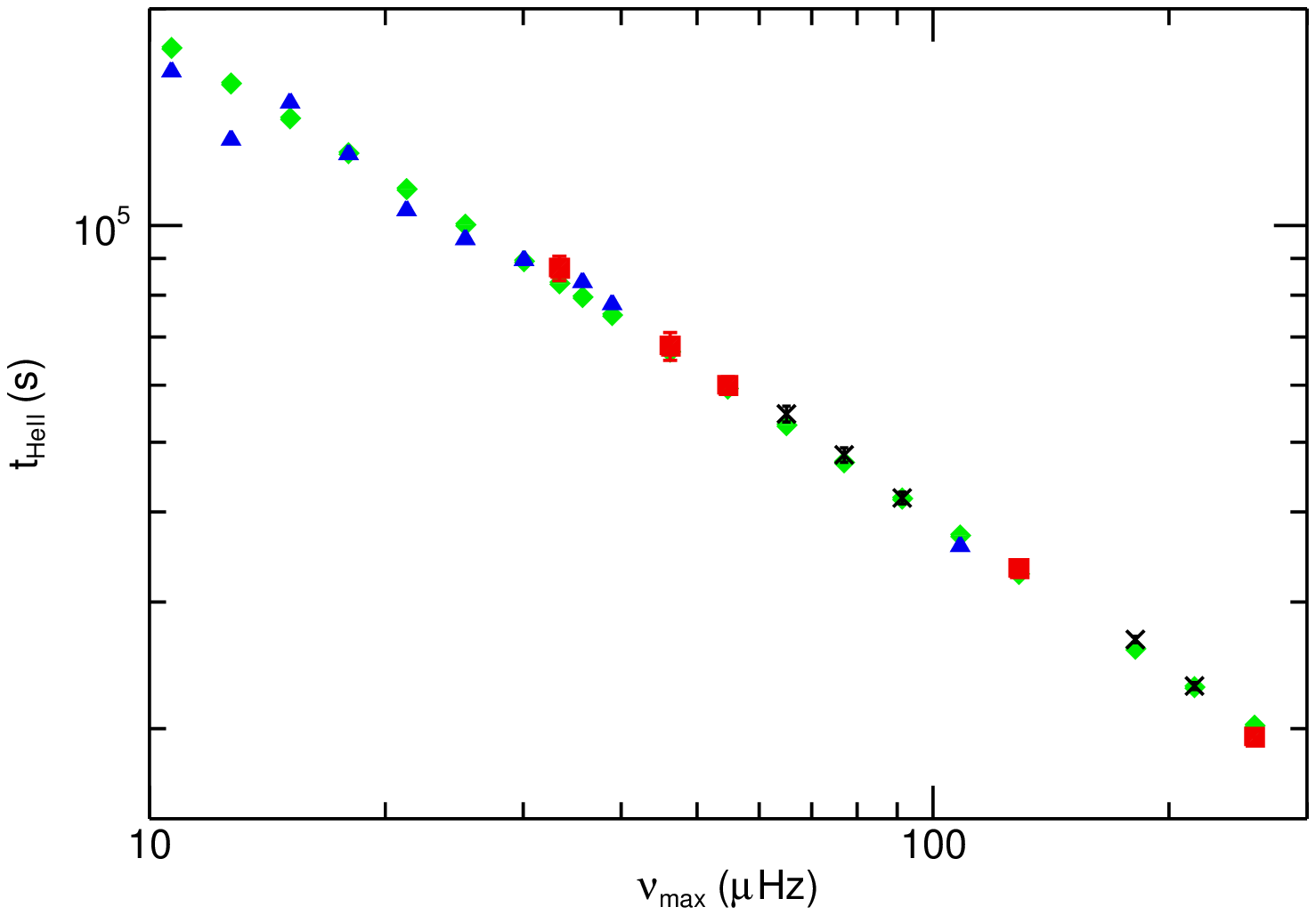}\\
  \includegraphics[width=0.4\textwidth, clip]{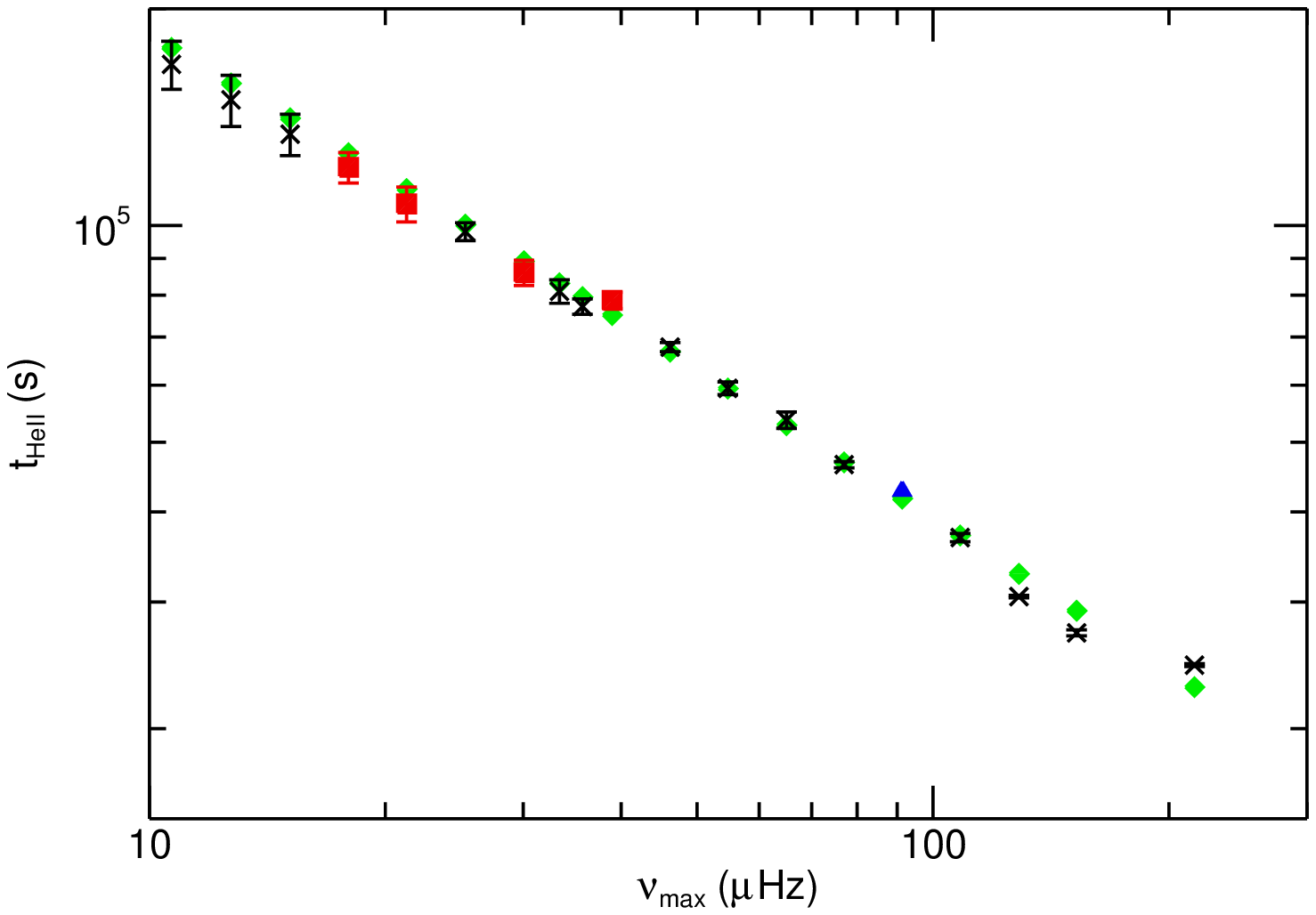}\\
  \caption{Acoustic radius of the second ionization zone of helium, $t_{\textrm{\scriptsize{HeII}}}$, as a function of $\nu_{\textrm{\scriptsize{max}}}$. Top panel: $l=0$ and 1 modes used. Middle panel: $l=0$ and 2 modes used. Bottom panel: $l=0$, 1, and 2 modes used. In all panels M1 RGB star models were used.}\label{figure[numax vs depth l]}
\end{figure}

\begin{figure}
    \centering
  \includegraphics[width=0.4\textwidth]{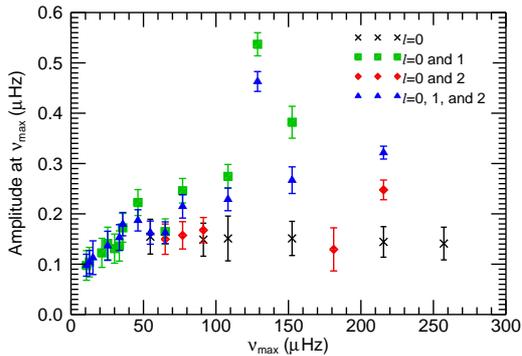}\\
  \caption{Amplitude of the signal from the helium second ionization code as a function of $\nu_{\textrm{\scriptsize{max}}}$. The different symbols represent the different combinations of modes that are used to fit the oscillatory component (see legend). Only fits that were classified as robust have been plotted. Here the results for the M1 models for RGB stars with $Y=0.278$ and $Z=0.020$ are plotted.}\label{figure[l012 amplitudes]}
\end{figure}

The effect of using $l=1$ modes in addition to $l=0$ modes is dependent on $\nu_{\textrm{\scriptsize{max}}}$ (see Fig. \ref{figure[numax vs depth l]}): first we notice that the models with the highest $\nu_{\textrm{\scriptsize{max}}}$ ($>70\,\rm\mu Hz$) either could not be fit or produced $t_{\textrm{\scriptsize{HeII}}}$ that were a long way from the model values. Fig. \ref{figure[l012 amplitudes]} compares the $A_{\textrm{\scriptsize{max}}}$ obtained when different combinations of harmonic degrees are used to fit the glitch. At high $\nu_{\textrm{\scriptsize{max}}}$ the amplitudes are overestimated when the $l=1$ modes are included. This occurs because the $l=1$ modes are mixed and so behave differently to the $l=0$ modes and so the oscillatory component exhibited by the $l=1$ modes has a different period and amplitude to the oscillatory component displayed by the $l=0$ modes (see Fig. \ref{figure[eg fit l]} for an example). Therefore, at high $\nu_{\textrm{\scriptsize{max}}}$ using $l=1$ modes could lead to an overestimation of the helium abundance of the star.

We note here that some of the fits that use mixed $l=1$ modes are nonetheless classified as robust and this may be misleading. Here the second differences of some of the $l=1$ modes are close to the second differences of the radial modes, however, because they are mixed they display a different periodicity and amplitude from the $l=0$ modes (see Fig. \ref{figure[eg fit l]}). The result is a relatively stable fit with an incorrect periodicity and amplitude. A comparison between the amplitudes and periodicities obtained when the $l=1$ modes are used and when only the $l=0$ modes are used may reveal that it is not appropriate to include the $l=1$ modes in these instances.

When $\nu_{\textrm{\scriptsize{max}}}<70\,\rm\mu Hz$ the fits are either robust or mostly robust and the differences between the model and fitted $t_{\textrm{\scriptsize{HeII}}}$ are less than $2\sigma$. Furthermore, the amplitudes agree with those obtained using $l=0$ modes only. This is particularly important at $\nu_{\textrm{\scriptsize{max}}}<40\,\rm\mu Hz$ where no robust fits were made when only radial modes were used. This behaviour is observed because as $\nu_{\textrm{\scriptsize{max}}}$ decreases the $l=1$ modes become more efficiently trapped in the acoustic cavity and so interact with the second ionization zone of helium in a similar manner to the radial modes, consequently aiding the fitting process. Being able to include $l=1$ modes improves the resolution of the second differences by approximately a factor of 2 since the separation of the $l=0$ and $l=1$ modes is approximately half the large separation. This is particularly important at low $\nu_{\textrm{\scriptsize{max}}}$, where the periodicity of the signal is similar to the large separation (and therefore the resolution of the second differences when only the $l=0$ modes are used). We note that although there are only a few models suitable for comparison it does appear that, as one might expect, the use of $l=1$ modes does reduce the size of the uncertainties associated with $A_{\textrm{\scriptsize{max}}}$. This could be important when attempting to discriminate between stars with different $Y$ (see Section \ref{section[model comparison]}).

When the $l=2$ modes are used as well as the $l=0$ modes the fit was improved in only six out of 21 models (see Fig. \ref{figure[numax vs depth l]}). The trapping of the $l=2$ modes in the acoustic cavity is efficient enough for the $l=0$ and $2$ modes to be affected by the glitch due to the second ionization zone of helium in a similar manner when $\nu_{\textrm{\scriptsize{max}}}<100\,\rm\mu Hz$. This is also evident in the amplitude of the oscillatory signal (see Fig. \ref{figure[l012 amplitudes]}): the agreement between the amplitudes obtained using $l=1$ modes and the amplitudes obtained using $l=0$ modes only breaks down for stars with $\nu_{\textrm{\scriptsize{max}}}>70\,\rm\mu Hz$. However, the amplitude obtained when $l=2$ modes are used appears to be similar to the amplitude obtained when only $l=0$ modes are used for stars with $\nu_{\textrm{\scriptsize{max}}}>70\,\rm\mu Hz$. Even when $\nu_{\textrm{\scriptsize{max}}}>100\,\rm\mu Hz$ some of the $l=2$ modes have second differences that are similar to those observed for the $l=0$ modes and consequently the fitting is marginally improved (see Fig. \ref{figure[eg fit l]}). When  $\nu_{\textrm{\scriptsize{max}}}<100\,\rm\mu Hz$, and the $l=2$ modes are efficiently trapped, any improvements in the fitted results are still marginal and, in the majority of cases, the improvement is not sufficient to make the fit robust. This is because the $l=0$ and 2 modes are only separated in frequency by a small amount and so, unlike when the $l=1$ modes are used and the resolution is significantly improved, inclusion of the $l=2$ modes only marginally improves the resolution of the second differences. Again, this is particularly important at low $\nu_{\textrm{\scriptsize{max}}}$ since it was the resolution of the second differences that prevented robust fits being made.

When $l=0$, 1, and 2 modes are used all of the fits with $\nu_{\textrm{\scriptsize{max}}}<80\,\rm\mu Hz$ are improved compared to the $l=0$ only fits. However, in terms of both $t_{\textrm{\scriptsize{HeII}}}$ and the amplitudes no significant gain is made by including the $l=2$ modes, i.e. the fitted results are similar to when just the $l=0$ and 1 modes are used.

Since the smooth component of equation (\ref{equation[fitted function]}) depends on the inertia of the modes (equation 6 in \citealt{Houdek2007}) we also investigated the effect of using different smooth components for modes of different $l$. However, we found that this had little or no effect on the fitted results (the change in both $t_{\textrm{\scriptsize{HeII}}}$ and the amplitude of the signal at $\nu_{\textrm{\scriptsize{max}}}$ was less than $1\sigma$).

Clearly care must be taken when including non-radial modes with regard to whether they are truly p modes or whether their frequency is significantly perturbed by gravity modes. We note that, although for these models there are enough higher order overtones to often be able to tell by eye whether they are behaving like p modes, this may not necessarily be the case for real data. However, if one is confident that the non-radial modes are behaving like p modes significant improvement can be obtained by using the $l=1$ non-radial modes in addition to the radial modes. Another potential way of improving the fitting procedure would be to include a larger number of radial modes, as we now discuss.

\subsection{The effect on the fitted parameters of the number of modes available}\label{section[number modes]}

Changing the limits on the range of modes used to fit the second differences so one extra high-frequency mode or one extra low-frequency mode was included improves the fit in just over half the models. The most notable improvements were observed at low $\nu_{\textrm{\scriptsize{max}}}$ as many more robust fits were made. For example, when using the original range in modes no robust fits were achieved when $\nu_{\textrm{\scriptsize{max}}}<40\,\rm\mu Hz$. However, four robust or mostly robust fits were achieved when one additional high-frequency mode was included and when one additional low-frequency mode was included. Across the entire range of $\nu_{\textrm{\scriptsize{max}}}$ considered here the values of $t_{\textrm{\scriptsize{HeII}}}$ obtained with the wider range were similar to those obtained with the original range, as were the sizes of the uncertainties. Similarly, when the fits were either robust or mostly robust, the differences in $A_{\textrm{\scriptsize{max}}}$ were significantly less than $1\sigma$. When equation (\ref{equation[FWHM gaussian]}) was used to determine the range of modes there are at most only seven second differences available to fit the oscillatory function. If this range were reduced further we would be unable to obtain robust fitted parameters for the majority of models under consideration. These results therefore suggest that the wider the range of modes we can use in the fits the better.

However, some of the underlying assumptions we have made restrict the range in $n$ for which a good fit can be made. By including a constant smooth component $(K)$ in equation (\ref{equation[fitted function]}), rather than a smooth component with higher order terms, we are assuming that the mode frequencies do not depart significantly from the asymptotic pattern. Over the limited range in frequency that we have considered so far this assumption is adequate. However, if we consider a much wider range of modes the assumption will break down and a higher order function would be required in addition to the periodic signal. However, since, for the majority of stars, we can only make use of the $l=0$ modes (see Section \ref{section[results diff l]}), we are unlikely to have enough modes to successfully fit a higher order smooth component in addition to the periodic signal.

\begin{figure}
\centering
  \includegraphics[width=0.4\textwidth, clip]{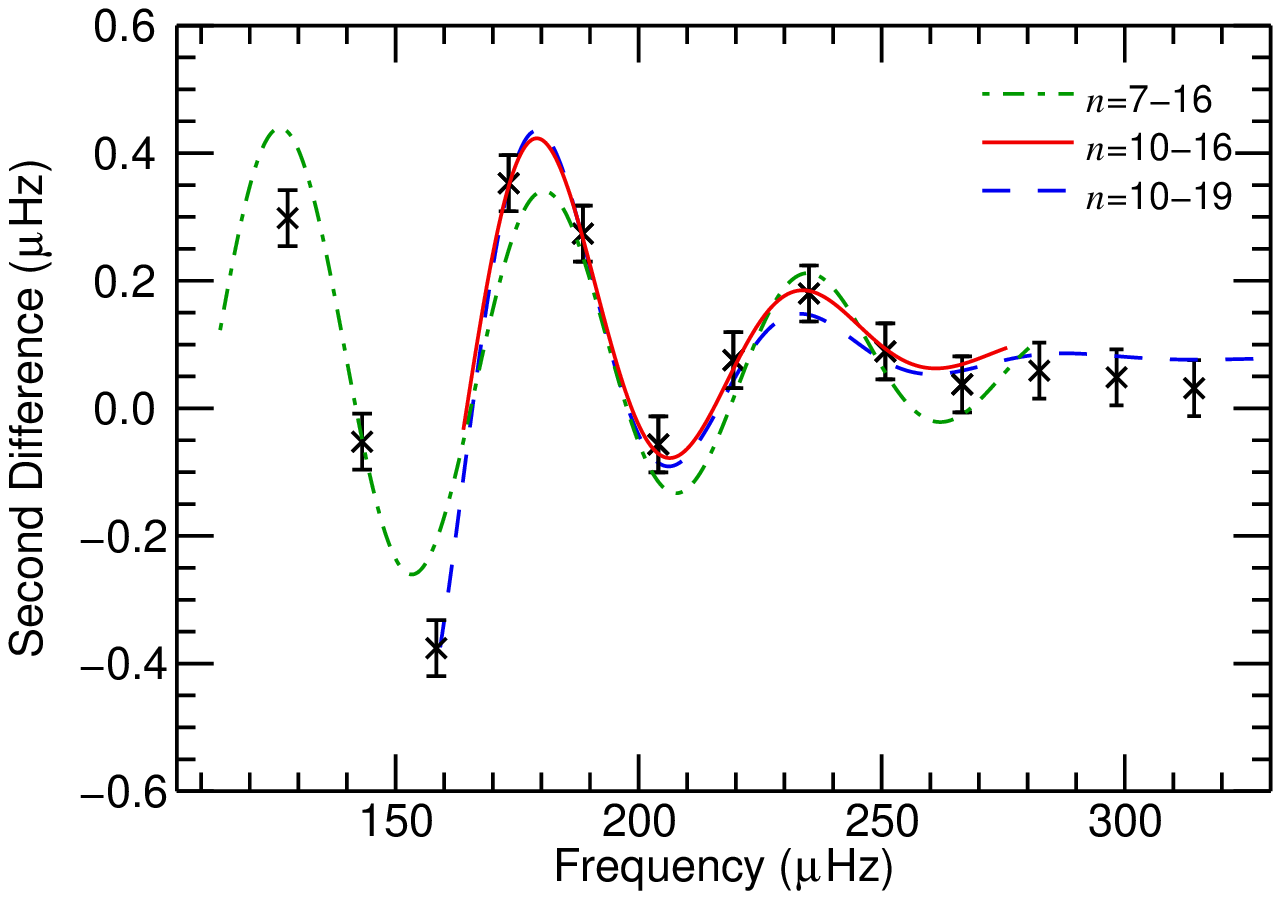}\\
  \includegraphics[width=0.4\textwidth, clip]{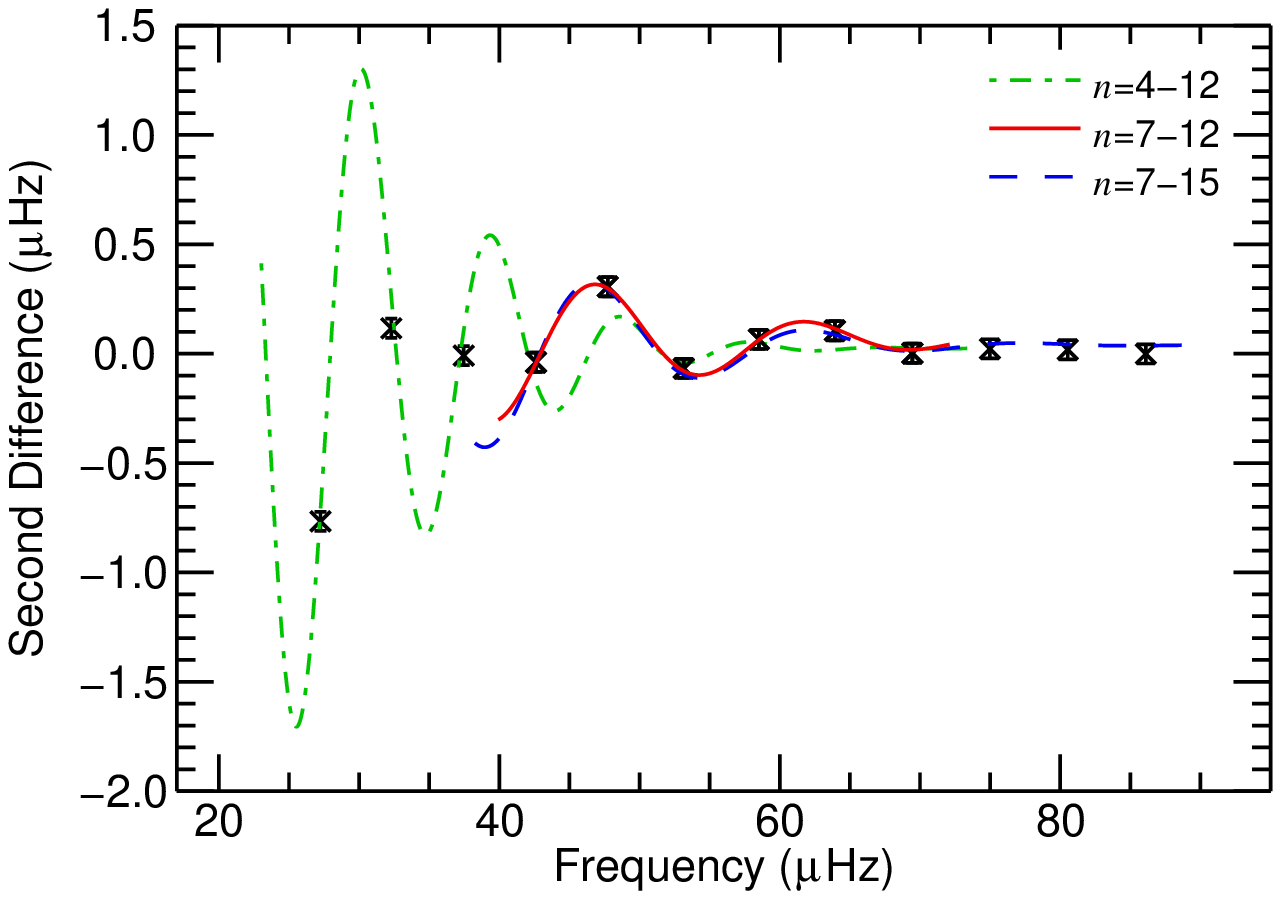}\\
  \caption{Functions fitted to the second differences when different ranges in overtones were used. The examples shown are for the same models as those shown in Fig. \ref{figure[eg fit]}.}\label{figure[eg fit diff n]}
\end{figure}

Fig. \ref{figure[eg fit diff n]} shows examples of the functions fitted to the second differences when the upper and lower boundaries on the range of overtones used to make the fit were changed. Changing the upper boundary had only a marginal effect on the fitted function with parameters, such as $t_{\textrm{\scriptsize{HeII}}}$ \textbf{and $A_{\textrm{\scriptsize{max}}}$}, changing by just a small fraction of the associated errors. Furthermore, the errors associated with the fitted parameters were reduced by only a marginal amount. Therefore, although it is beneficial to have additional high-frequency overtones the effect of the extra overtones on the accuracy and precision of the fits is marginal.

When the lower limit on the range of modes used to fit the second differences was decreased (and so more low-frequency modes were included) the values of the fitted parameters changed significantly. Fig. \ref{figure[eg fit diff n]} shows that when lower frequency modes were used the resultant fitted function represented the modelled second differences well at low frequencies, but less well at high frequencies. This is because the smooth component is no longer approximately constant and so the period of the signal appears to change with frequency. Care must therefore be taken when including low-frequency modes. However, the amplitude of the signal increases exponentially with decreasing frequency and so a balance must be made between using overtones where the amplitude of the signal is large enough to fit accurately and precisely and introducing systematic errors by straying outside the asymptotic regime. We suggest that the most pragmatic way to proceed would be to test the stability of the fitted parameters when the lower limit on the range of modes used is varied.

\subsection{The effect on the fitted parameters of the size of the uncertainties on the observed
frequencies}\label{section[results diff errors]}

As previously stated the uncertainties of the model frequencies have been added artificially and we have scaled the uncertainties so as to be approximately appropriate for 4\,yr of \textit{Kepler} data. However, our approach is in many ways simplistic as it does not take into account factors such as the signal-to-noise ratio and lifetimes of the oscillations. Therefore, we tested the quality of the fits when the size of the uncertainties on the second difference of the mode frequencies was both increased to be 25\,percent larger and decreased to be 25\,percent smaller than the original uncertainties.

\subsubsection{Comparison of fits when uncertainties were 25\,percent smaller}
When the size of the error bars on the second differences was scaled so as to be 25\,per cent smaller the values of the fitted $t_{\textrm{\scriptsize{HeII}}}$ and $A_{\textrm{\scriptsize{max}}}$ changed by less than 0.001\,per cent compared to the fitted $t_{\textrm{\scriptsize{HeII}}}$ \textbf{and $A_{\textrm{\scriptsize{max}}}$} obtained using the original uncertainties. This is significantly less than the error bars in the fitted parameters (see Table \ref{table[errors]}). We define $\sigma_{\textrm{\scriptsize{O}}}$ as the uncertainties on the fitted parameters using the original uncertainties on $\Delta_2\nu_{n,l}$, and $\sigma_{0.75\textrm{\scriptsize{O}}}$ as the uncertainties on the fitted parameters when the uncertainties on $\Delta_2\nu_{n,l}$ were 25\,percent smaller. As one might expect we find that $\sigma_{0.75\textrm{\scriptsize{O}}}$ are approximately 25\,percent smaller than $\sigma_{\textrm{\scriptsize{O}}}$ (21\,percent for  $A_{\textrm{\scriptsize{max}}}$, see Table \ref{table[errors]}). Fits were obtained for the same models regardless of the size of the uncertainties and the robustness of the fits was either the same (12 out of 17 models) or improved (five out of 17 models).

\begin{table*}\caption{Mean uncertainties associated with the fitted $t_{\textrm{\scriptsize{HeII}}}$ when the uncertainties on the second differences were both increased and decreased.}\label{table[errors]}
\begin{tabular}{|c|c|c|c|c|c|}
  \hline
   & \multicolumn{3}{|c|}{Mean uncertainty as percentage of parameter} & \multicolumn{2}{c|}{Ratio} \\
  \hline
  Parameter & $\sigma_{\textrm{\scriptsize{O}}}$ & $\sigma_{0.75\textrm{\scriptsize{O}}}$ & $\sigma_{1.25\textrm{\scriptsize{O}}}$ & $\sigma_{0.75\textrm{\scriptsize{O}}}/\sigma_{\textrm{\scriptsize{O}}}$ & $\sigma_{1.25\textrm{\scriptsize{O}}}/\sigma_{\textrm{\scriptsize{O}}}$ \\
  \hline
  $t_{\textrm{\scriptsize{HeII}}}$ & 2.53 & 1.93 & 3.81 & 0.75 & 1.24 \\
  $A_{\textrm{\scriptsize{max}}}$ & 23.4 & 19.9 & 28.1 & 0.79 & 1.20 \\
  \hline
\end{tabular}
\end{table*}

\subsubsection{Comparison of fits when uncertainties were 25\,percent larger}
We define $\sigma_{1.25\textrm{\scriptsize{O}}}$ as the uncertainties on the fitted parameters when the uncertainties on $\Delta_2\nu_{n,l}$ were increased by 25\,percent. The values of the fitted $t_{\textrm{\scriptsize{HeII}}}$ and $A_{\textrm{\scriptsize{max}}}$ changed by less than $0.01$\,per cent compared to the values of the parameters found using the original uncertainties on $\Delta_2\nu_{n,l}$. Again as one might expect the average value of $\sigma_{1.25\textrm{\scriptsize{O}}}$ were approximately 25\,percent larger than $\sigma_{\textrm{\scriptsize{O}}}$ (20\,percent for $A_{\textrm{\scriptsize{max}}}$, see Table \ref{table[errors]}). Fits were obtained for the same set of models and the robustness of the fit was improved in two models, remained the same for 10 models and for five models the fits were deemed to be less robust.

\begin{figure}
\centering
  \includegraphics[width=0.4\textwidth, clip]{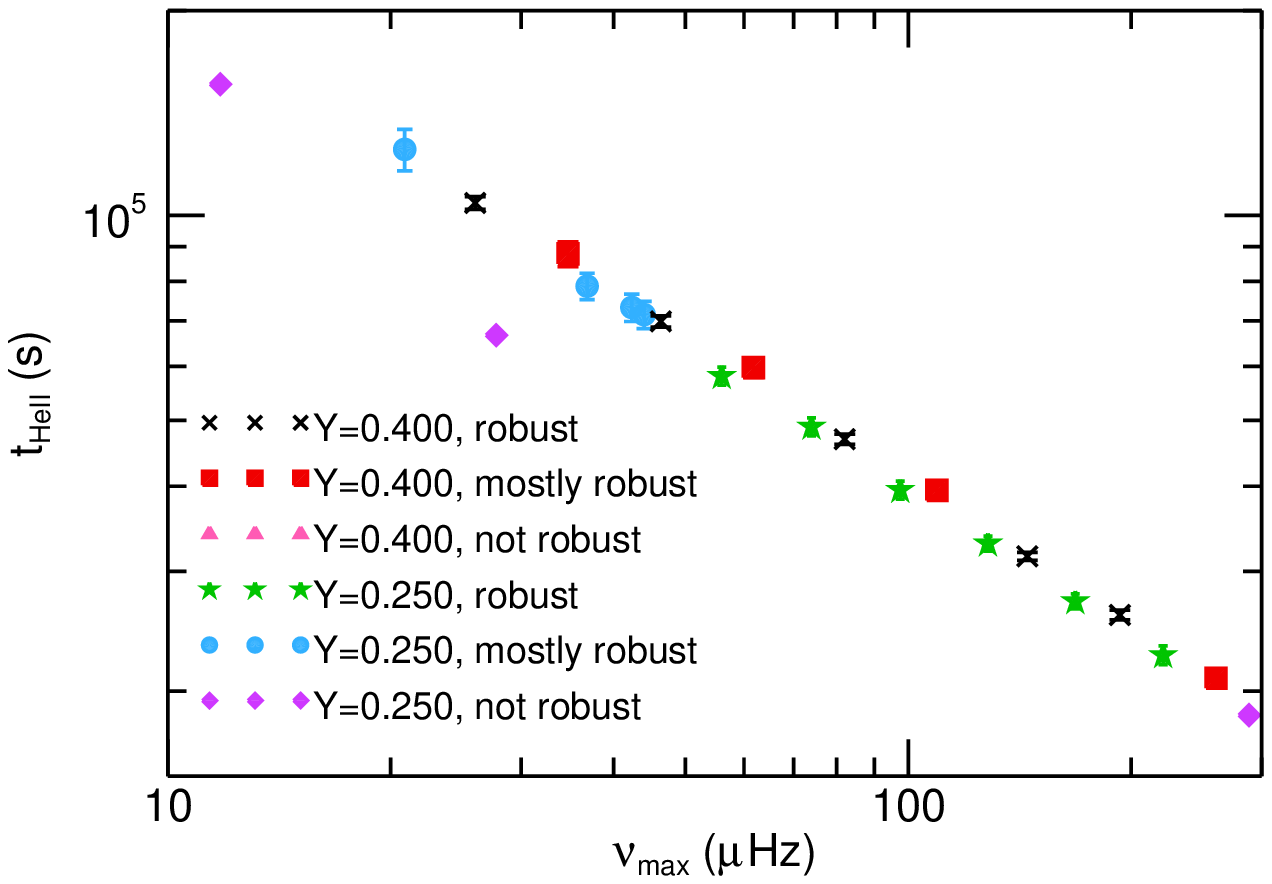}\\
  \includegraphics[width=0.4\textwidth, clip]{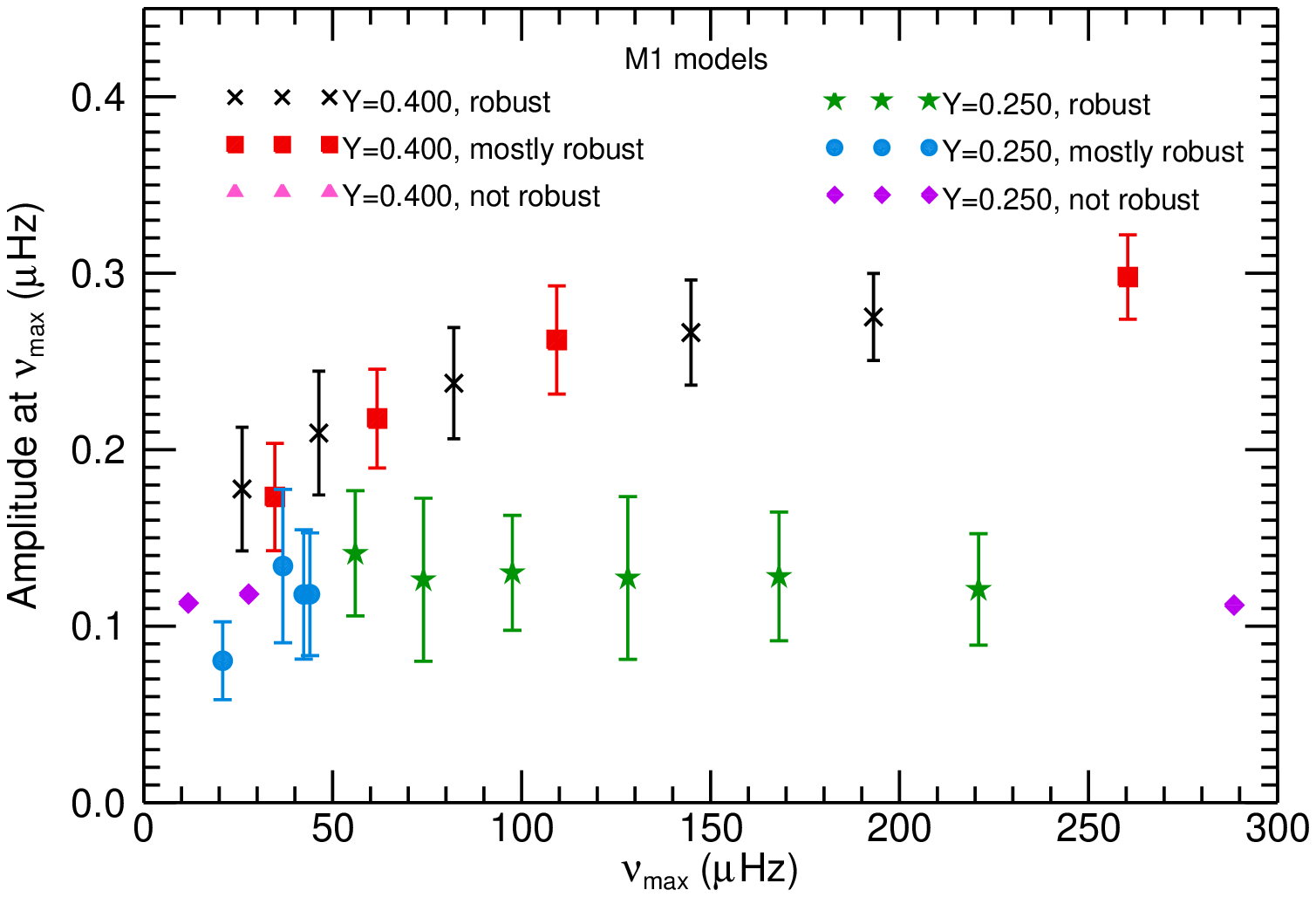}\\
  \caption{Top panel: acoustic radius of the second ionization zone of helium, $(t_{\textrm{\scriptsize{HeII}}})$, as a function of $\nu_{\textrm{\scriptsize{max}}}$. Bottom panel: amplitude of the signal from the helium secondionization code as a function of $\nu_{\textrm{\scriptsize{max}}}$. Each panel plots results for the M1 RGB star models with $Y=0.250$ and $Y=0.400$ (see legend).}\label{figure[numax vs depth extreme]}
\end{figure}

In conclusion, as one might expect, the better the constraints on the second differences, and therefore on the mode frequencies, the better the constraints on the fitted parameters. We recall that the size of the original uncertainties was proportional to the inverse of the square root of the length of the time series (see Section \ref{section[method]}). Since the size of the uncertainties on the parameters obtained from the fit is directly proportional to the uncertainties on the second differences the results of this section imply that the longer we can make asteroseismic observations of a star the tighter constraints we can place on the parameters of the second ionization zone of helium. All of the results outlined in this section have been obtained using the M1 models with $Y=0.278$. We now discuss whether these results are specific to these models.

\section{Comparison of different models}\label{section[model comparison]}

\subsection{M1 models with well-separated $Y$}

We begin by considering $t_{\textrm{\scriptsize{HeII}}}$ \textbf{and $A_{\textrm{\scriptsize{max}}}$} obtained from models generated with the same code as above (the M1 models) but with different initial helium abundances: the helium abundances considered were $Y=0.250$ and $Y=0.400$.

Fig. \ref{figure[numax vs depth extreme]} shows the variation in $t_{\textrm{\scriptsize{HeII}}}$ \textbf{and $A_{\textrm{\scriptsize{max}}}$} with $\nu_{\textrm{\scriptsize{max}}}$ for two different values of $Y$. For the robust fits, the gradient of the variation in $t_{\textrm{\scriptsize{HeII}}}$ with respect to $\nu_{\textrm{\scriptsize{max}}}$ is the same for both sets of models. For models with approximately the same $\nu_{\textrm{\scriptsize{max}}}$, the different helium abundances mean that the determined $t_{\textrm{\scriptsize{HeII}}}$ are offset since the second ionization zone of helium is closer to the surface of the star in the models with a higher initial helium abundance (or $t_{\textrm{\scriptsize{HeII}}}$ is larger when $Y=0.400$). However, this offset is similar in magnitude to the uncertainties associated with the fitted $t_{\textrm{\scriptsize{HeII}}}$, meaning that it would be difficult to use the obtained values of $t_{\textrm{\scriptsize{HeII}}}$ to distinguish between stars with different initial helium abundances. The size of the uncertainties associated with the fitted values of $t_{\textrm{\scriptsize{HeII}}}$ is smaller for the $Y=0.400$ models than those associated with the $Y=0.250$ models. This is because the higher abundance means that the amplitude of the signal is larger, thereby making the signal easier to fit. This is evident in Fig. \ref{figure[numax vs depth extreme]} which shows that robust fits are obtained at lower frequencies when $Y=0.400$ than when $Y=0.250$ (and $Y=0.278$).

Perhaps more useful is the bottom panel of Fig. \ref{figure[numax vs depth extreme]} which shows the variation in $A_{\textrm{\scriptsize{max}}}$ with $\nu_{\textrm{\scriptsize{max}}}$ for the two extreme values of $Y$. When $\nu_{\textrm{\scriptsize{max}}}\gtrsim50\,\rm\mu Hz$ the amplitudes follow two distinct tracks for the different helium abundances. This implies that, particularly when $\nu_{\textrm{\scriptsize{max}}}\gtrsim100\,\rm\mu Hz$, we would be able to differentiate between stars with these two well separated abundances.

\begin{figure*}
\centering
  \includegraphics[width=0.4\textwidth, clip]{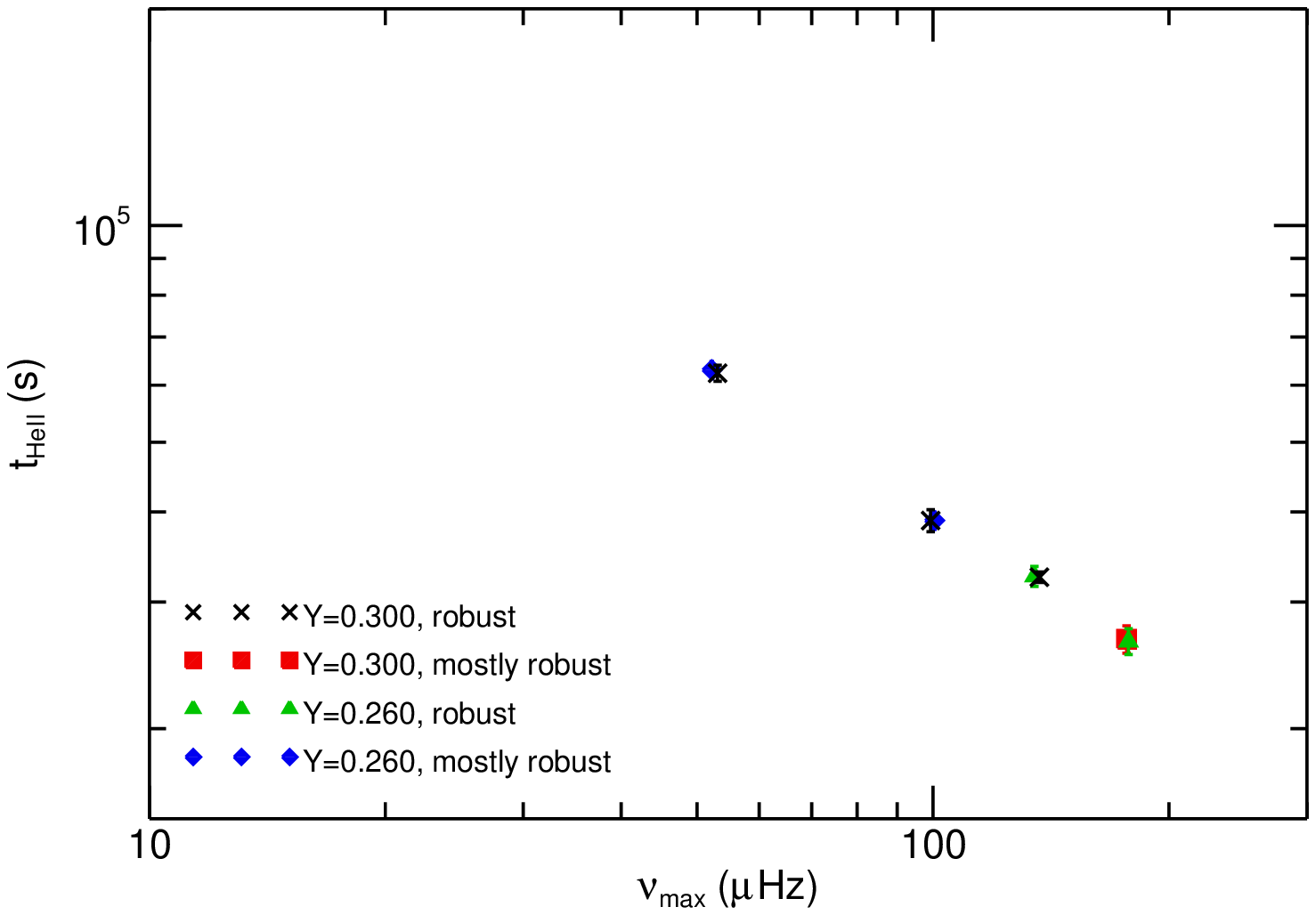}
  \includegraphics[width=0.4\textwidth, clip]{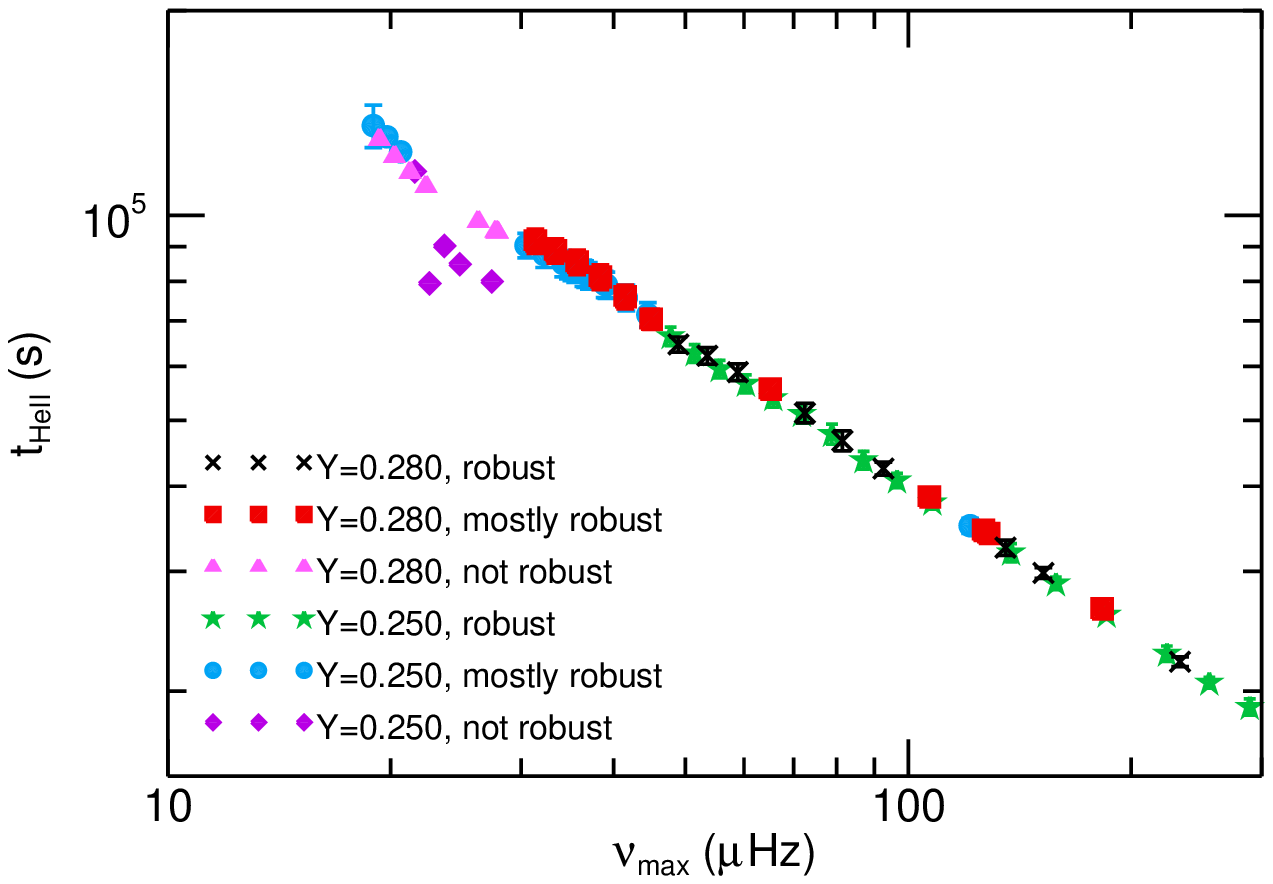}\\
  \includegraphics[width=0.4\textwidth, clip]{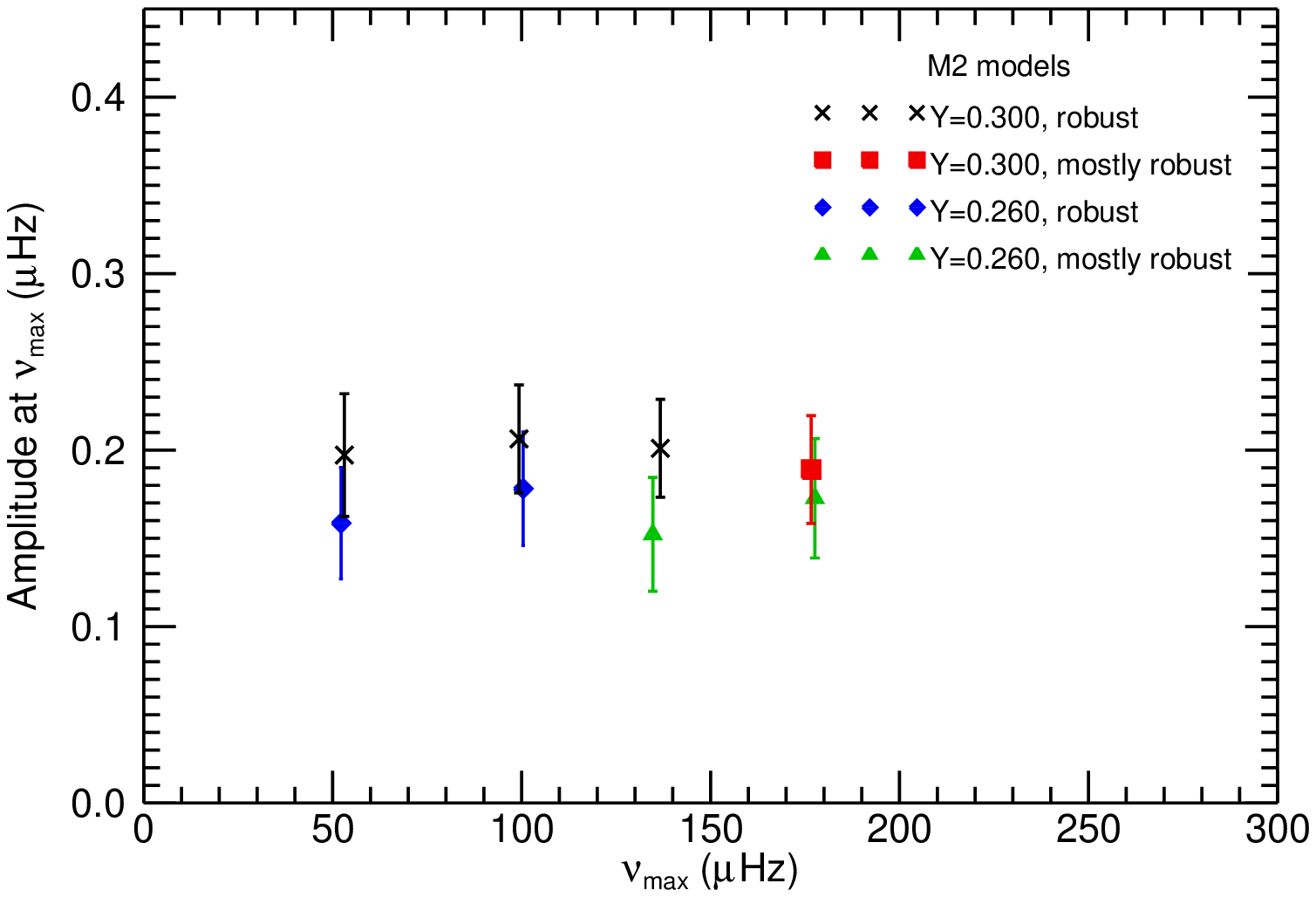}
  \includegraphics[width=0.4\textwidth, clip]{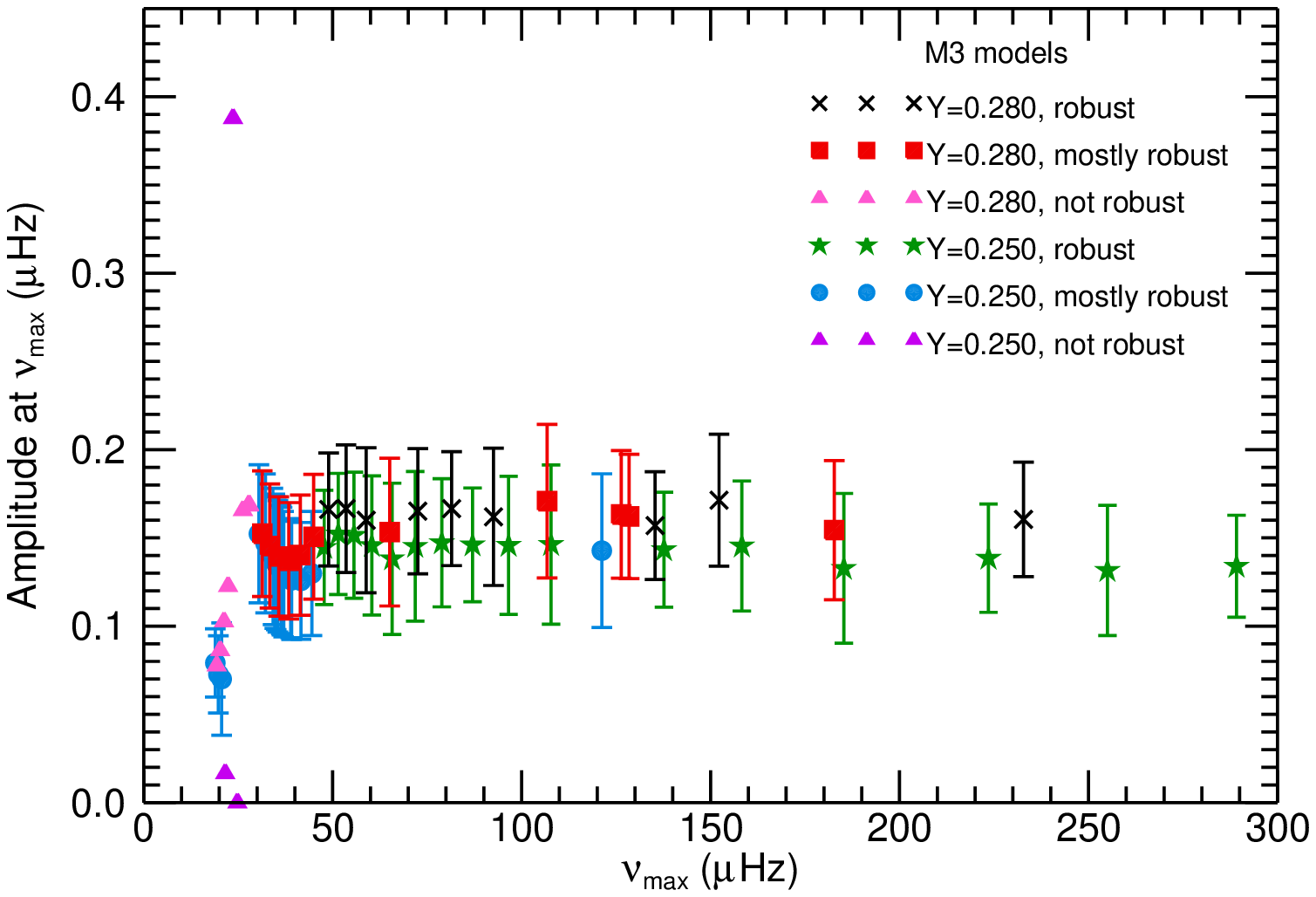}\\
  \caption{Top left: acoustic radius of the second ionization zone of helium $(t_{\textrm{\scriptsize{HeII}}})$  as a function of $\nu_{\textrm{\scriptsize{max}}}$ for M2 RGB star models. Top right: acoustic radius of the second ionization zone of helium $(t_{\textrm{\scriptsize{HeII}}})$  as a function of $\nu_{\textrm{\scriptsize{max}}}$ for M3 RGB star models. Bottom left: amplitude of the acoustic glitch at $\nu_{\textrm{\scriptsize{max}}}$ for the same models as plotted in the top left-hand panel of this figure. Bottom right: amplitude of the acoustic glitch at $\nu_{\textrm{\scriptsize{max}}}$ for the same models as plotted in the top right-hand panel of this figure.}\label{figure[numax vs depth other codes]}
\end{figure*}

\subsection{M2 and M3 models}

The left-hand panels of Fig. \ref{figure[numax vs depth other codes]} show results obtained when the frequencies were generated using the M2 code. In this instance the masses of the stars were $1.6\,\rm M_\odot$ (i.e. slightly more massive than the M1 models). Two different helium abundances are shown: one slightly higher than those used to generate the stars shown in Fig. \ref{figure[numax vs depth]} ($Y=0.300$) and one slightly lower ($Y=0.260$). Despite the difference in $Y$ and mass, and the different model generation codes the results are in good agreement with those seen in Figs. \ref{figure[numax vs depth]} and \ref{figure[amplitudes]}.

In the right-hand panels of Fig. \ref{figure[numax vs depth other codes]} (which uses the M3 models) the stars are slightly more massive than those examined in Figs. \ref{figure[numax vs depth]} and \ref{figure[amplitudes]} but the helium abundance and metallicity are the same. Again the results are in good agreement with those observed in Figs. \ref{figure[numax vs depth]} and \ref{figure[amplitudes]} despite the different model generation code. We also observed the same deterioration in the robustness of the fits at $\nu_{\textrm{\scriptsize{max}}}\lesssim40\,\rm\mu Hz$. Fig. \ref{figure[numax vs depth other codes]} therefore implies that the results described in Section \ref{section[results]} are not model specific.

Although large differences $(\delta Y~0.15)$ in the initial helium abundances of red giants may be observed in globular clusters and a small fraction of halo stars \citep[e.g.][]{DAntona2004, Piotto2005, Piotto2007, Sollima2005, Lee2005}, in the galactic disc the difference in $Y$ is likely to be significantly less than this \citep[e.g.][]{Chiosi1982, Bressan2012}. In the bottom left-hand panel of Fig. \ref{figure[numax vs depth other codes]} we consider a difference in $Y$ of 0.04. Although a difference in the amplitude is observed this difference is always less than $1.5\sigma$, indicating that, with the errors assumed here, one would not be able to discriminate between stars with such similar $Y$.

We remind the reader that the size of the error bars on the second differences, and therefore on the fitted parameters, has been artificially added. In each panel of Fig. \ref{figure[numax vs depth other codes]} the uncertainties were appropriate for approximately 1460\,d of data. However, in Section \ref{section[results diff errors]} we found that the size of the errors on $t_{\textrm{\scriptsize{HeII}}}$ and $A_{\textrm{\scriptsize{max}}}$ approximately scaled depending on the size of the errors on $\Delta_2\nu_{n,l}$. It is therefore reasonable to ask how long a data set is required before we can differentiate between stars with $Y=0.260$ and $0.300$?

Using a simple scaling where the size of the uncertainties on the amplitude at $\nu_{\textrm{\scriptsize{max}}}$ is proportional to the size of the uncertainties associated with the original frequencies, the results shown in the bottom left-hand panel of Fig. \ref{figure[numax vs depth other codes]} indicate that the difference between the amplitudes will be greater than $3\sigma$ if the uncertainties on the original frequencies are less than $0.005\,\rm\mu Hz$. As mentioned in Section \ref{section[method]} the size of the uncertainties on the frequencies is proportional to the inverse square root of the length of the observations. Using this simple scaling this level of precision would require over 60\,yr of observations.

\begin{figure}
  \includegraphics[width=0.4\textwidth, clip]{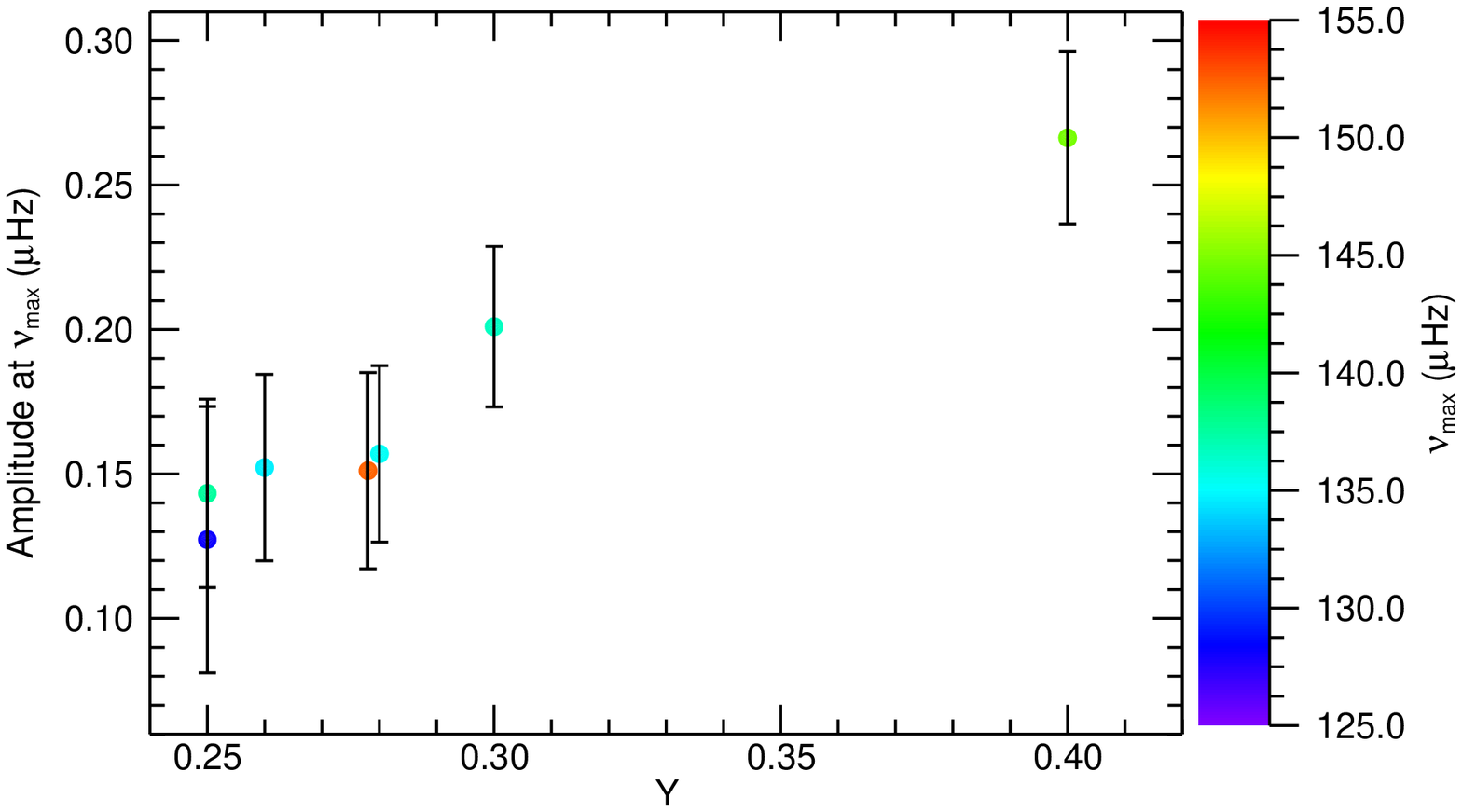}\\
  \includegraphics[width=0.4\textwidth, clip]{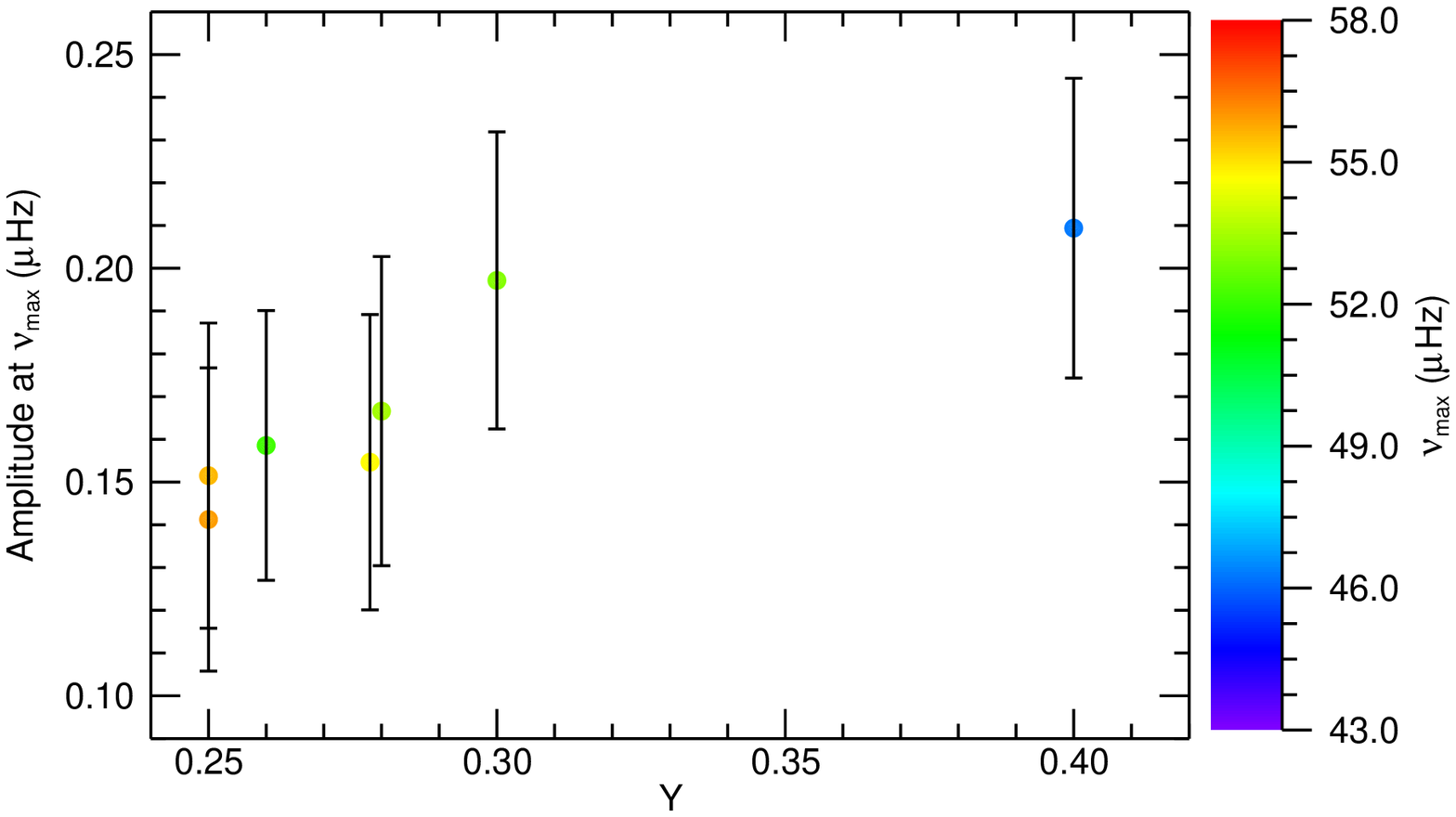}\\
  \caption{Amplitude of the signal from the helium second ionization code as a function of $Y$. In each figure the colour of the symbol was determined by $\nu_{\textrm{\scriptsize{max}}}$. One fit was classified as mostly robust (in the top panel) but all other fits were classified as robust. M1, M2, and M3 models have been used. Each individual panel shows models with similar $\nu_{\textrm{\scriptsize{max}}}$: in the top panel $\nu_{\textrm{\scriptsize{max}}}\approx140\,\rm\mu Hz$, whereas in the bottom panel $\nu_{\textrm{\scriptsize{max}}}\approx50\,\rm\mu Hz$. All models are for RGB stars. We note here that all of the amplitudes have been obtained using $l=0$ modes only as for the models used to produce the top panel the non-radial modes are likely to be mixed.}\label{figure[y vs amplitudes]}
\end{figure}

Fig. \ref{figure[y vs amplitudes]} further demonstrates that there is clearly a trend for the amplitude of the signal at $\nu_{\textrm{\scriptsize{max}}}$ to increase with $Y$ when $\nu_{\textrm{\scriptsize{max}}}$ is relatively large. However, at lower $\nu_{\textrm{\scriptsize{max}}}$ this trend is noticeably less pronounced.

\subsection{Clump stars}

\begin{figure*}
  \includegraphics[width=0.42\textwidth, clip]{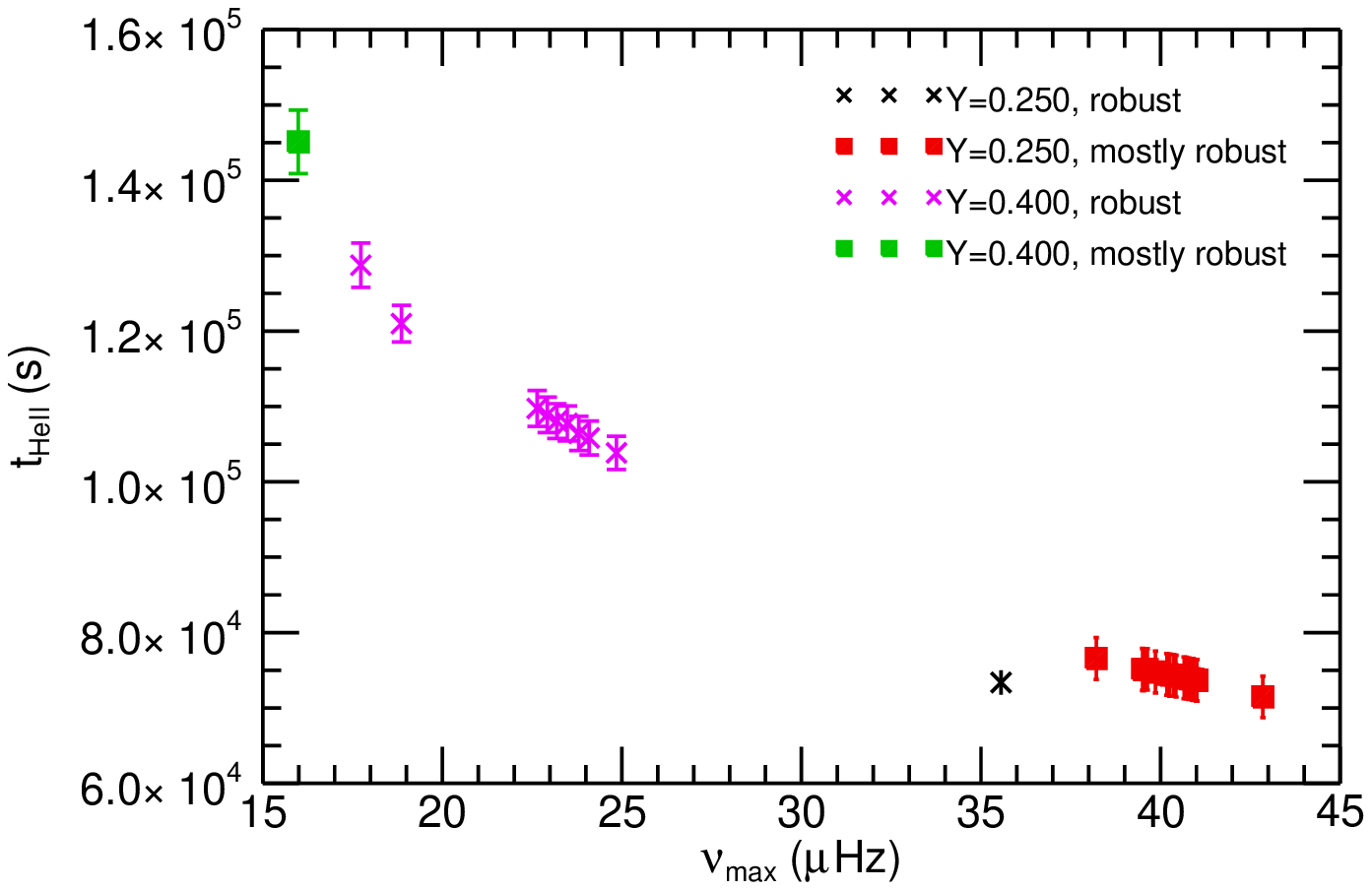}
  \includegraphics[width=0.42\textwidth, clip]{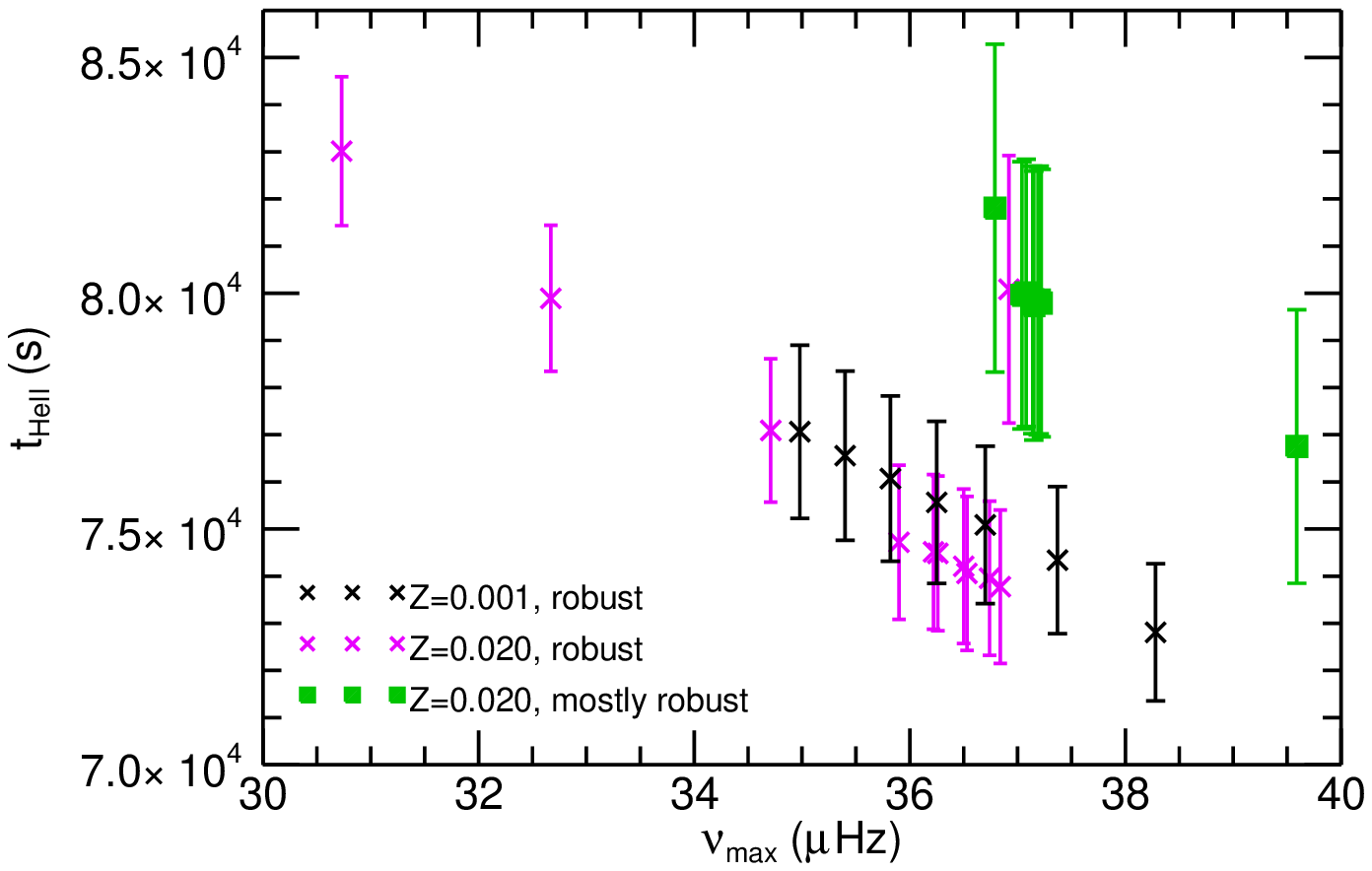}\\
  \hspace{0.4cm}\includegraphics[width=0.4\textwidth, clip]{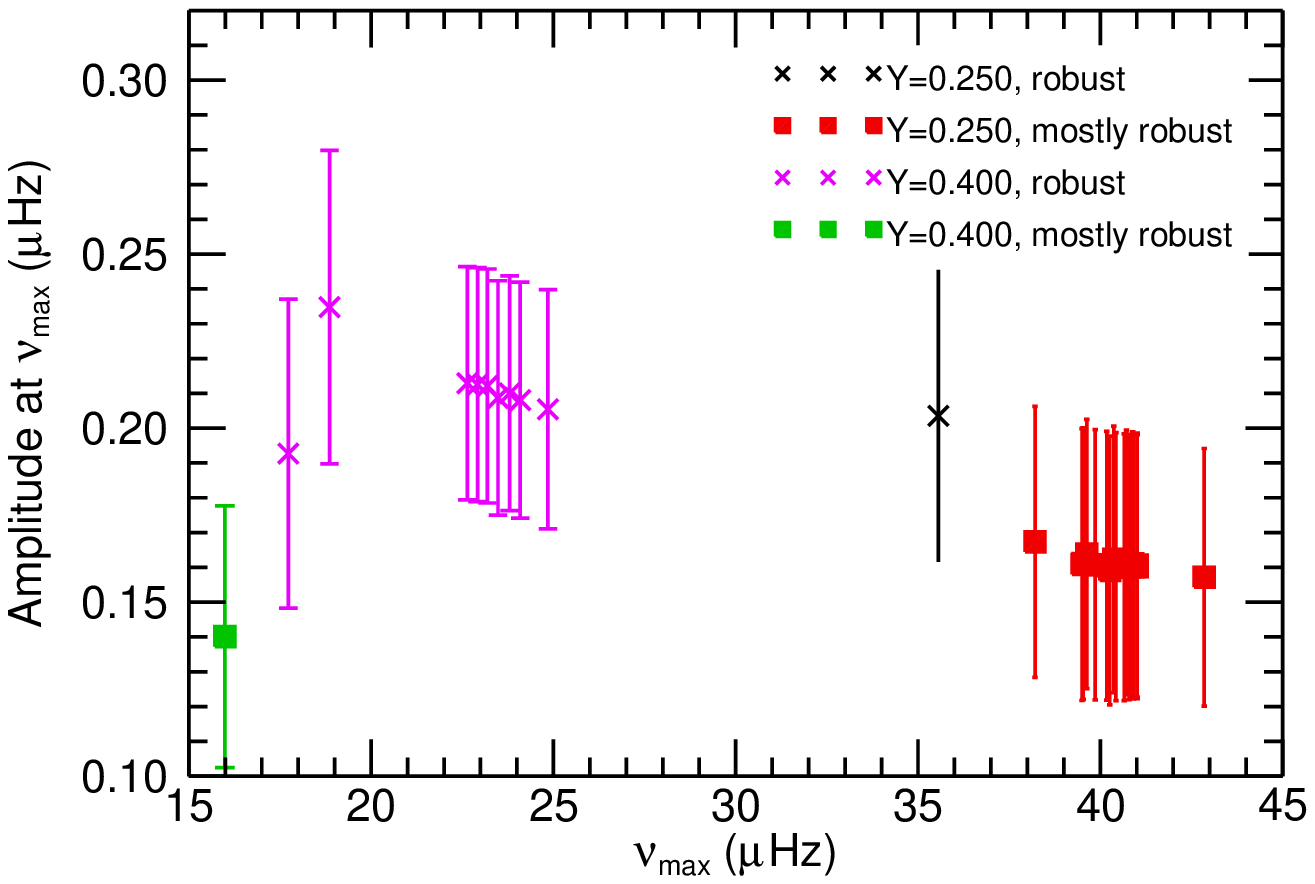}\hspace{0.4cm}
  \includegraphics[width=0.4\textwidth, clip]{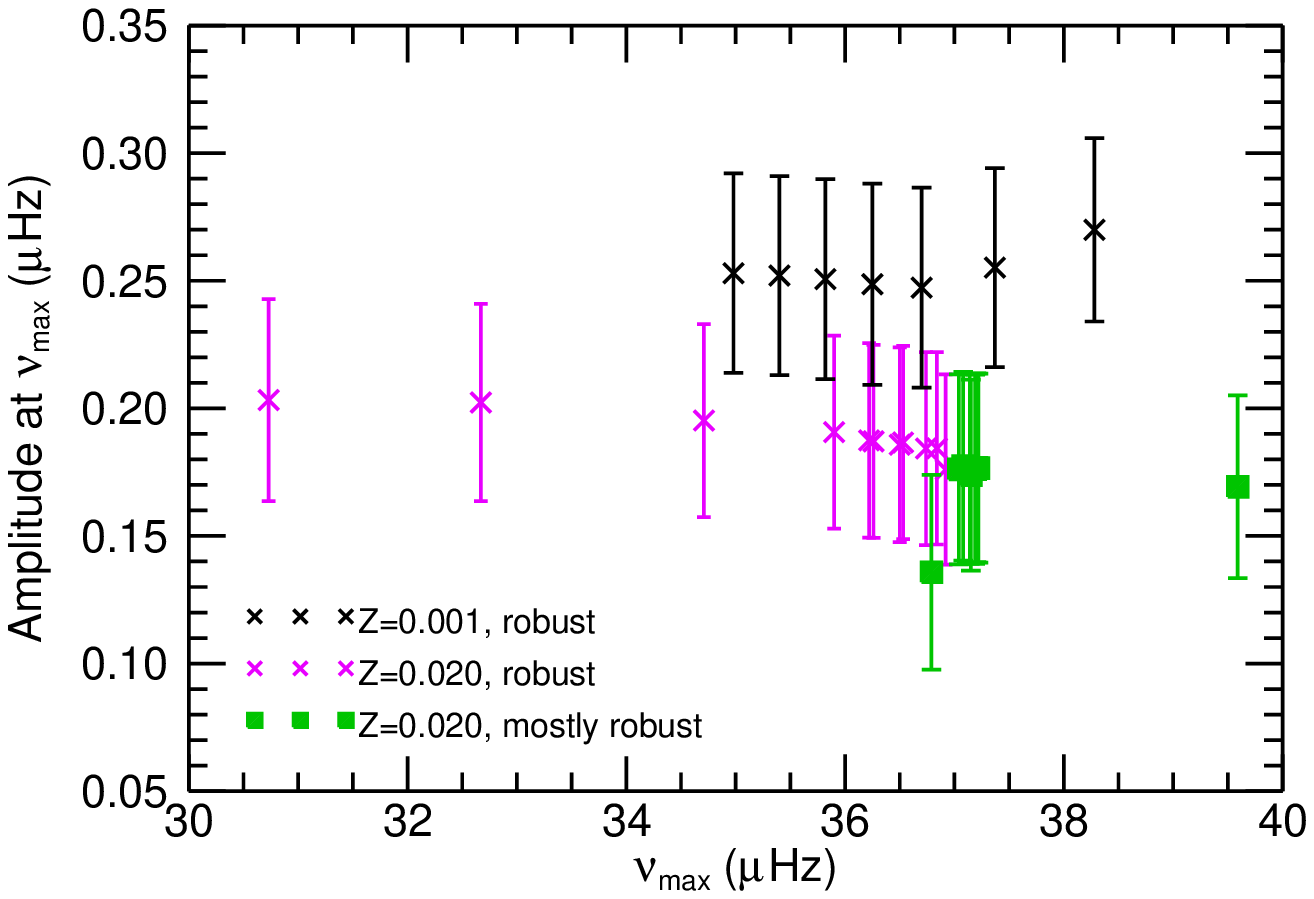}\\
  \caption{Top left: acoustic radius of the helium second ionization zone, $t_{\textrm{\scriptsize{HeII}}}$, as a function of $\nu_{\textrm{\scriptsize{max}}}$ for clump stars with two different initial helium abundances, $Y$ (see legend). M1 models with $Z=0.020$ were used. Top right: acoustic radius of the helium second ionization zone, $t_{\textrm{\scriptsize{HeII}}}$, as a function of $\nu_{\textrm{\scriptsize{max}}}$ for clump stars with two different initial heavy element abundances, $Z$ (see legend). M1 models with $Y=0.278$ were used. Bottom left: amplitude of the acoustic glitch at $\nu_{\textrm{\scriptsize{max}}}$ for the same models as plotted in the top left-hand panel of this figure. Bottom right: amplitude of the acoustic glitch at $\nu_{\textrm{\scriptsize{max}}}$ for the same models as plotted in the top right-hand panel of this figure. }\label{figure[clump]}
\end{figure*}

Fig. \ref{figure[clump]} shows the observed amplitude of the signal and $t_{\textrm{\scriptsize{HeII}}}$ obtained from the frequencies for clump stars when the models had different $Y$ and $Z$. The M1 models with a mass of $1.5\,M_\odot$ were used. Definite trends in both $t_{\textrm{\scriptsize{HeII}}}$ and the amplitude of the glitch at $\nu_{\textrm{\scriptsize{max}}}$ are observed. 
Models computed with significantly different $Y$ go through the clump phase at different radii, this is reflected in the different $\nu_{\textrm{\scriptsize{max}}}$. However, it is not possible to discriminate between stars with different $Z$.

We note that for model stars with approximately the same $\nu_{\textrm{\scriptsize{max}}}$, the difference between the fitted $t_{\textrm{\scriptsize{HeII}}}$ and amplitude at $\nu_{\textrm{\scriptsize{max}}}$ obtained for clump stars and RGB stars is of the order of $1\sigma$. Therefore it is not possible to discriminate between the two classes of stars based solely on an analysis of glitch caused by the second ionization zone of helium. \textbf{This is largely expected} since the layers where helium ionization takes place are largely determined by conditions at the surface, which are similar acoustically in RGB and clump stars.

\section{Discussion}\label{section[discussion]}

We have successfully used second differences of p-mode frequencies to gain information about the second ionization zone of helium for a wide range of model stars. Our methodology, particularly when $\nu_{\textrm{\scriptsize{max}}}>40\,\rm\mu Hz$, has been shown to be both robust and accurate. However, this study has highlighted the fact that comparing results obtained from the model-produced frequencies with those obtained directly from the models is not straightforward. Comparisons between results obtained from actual observed data and models would be even more uncertain. Comparisons of the acoustic depth of the ionization zone appear to be inconsistent raising questions over the treatment of the near-surface effects and even the definition of the acoustic surface. To avoid these uncertainties we have instead compared estimates of the acoustic radius of the second ionization zone of helium, $t_{\textrm{\scriptsize{HeII}}}$. However, even this is not straightforward: our results indicate that the acoustic radius of the glitch obtained by fitting frequencies is consistent with that of a local maximum in $\gamma_1$, rather than the local depression. We note here that other authors have also found discrepancies between $\tau_{\textrm{\scriptsize{HeII}}}$ defined in the models as the local minimum in $\gamma_1$ and those obtained from the fit \citep{Houdek2007, Mazumdar2014}.

The signature of the glitch in $\Delta_2\nu_{n,l}$ was difficult to fit at low $\nu_{\textrm{\scriptsize{max}}}$ ($<40\,\rm\mu Hz$) because the number of overtones was low and the periodicity of the glitch was similar in magnitude to the resolution of the second differences. Including higher order modes, particularly $l=1$ modes, did, in general, aid the fitting process but only when the modes were not mixed. Mixed modes are influenced by the glitch in a different manner to p modes and so the signature observed in the $\Delta_2\nu_{n,l}$ is different. Importantly, though, we found that for RGB stars the $l=1$ modes did behave like p modes at low $\nu_{\textrm{\scriptsize{max}}}$, where our methodology struggled to produce robust fits using the $l=0$ modes alone. However, we note that this is unlikely to be true for clump stars with the same $\nu_{\textrm{\scriptsize{max}}}$.

One might naively think that simply including more $l=0$ modes in the analysis, particularly at low frequencies where the amplitude of the signal is largest, might improve the quality of the fitted results. However, care must be taken not to stray outside the asymptotic regime as then the fitted function described by equation (\ref{equation[fitted function]}) becomes inappropriate: A higher order background function is required instead of the simple constant, $K$.

The acoustic radius is not the only parameter that can be obtained by fitting the signature of the acoustic glitch. The amplitude of the envelope of the signature at $\nu_{\textrm{\scriptsize{max}}}$ is correlated with the initial helium abundance of the star, $Y$. We note, however, that the amplitude at $\nu_{\textrm{\scriptsize{max}}}$ is not a straight measure of $Y$. Whether the amplitude at $\nu_{\textrm{\scriptsize{max}}}$ can be used to discriminate between stars of different $Y$ depends on the uncertainties associated with the p-mode frequencies. For the majority of this paper we have used uncertainties typical of 1460\,d of data (or $0.02\,\rm\mu Hz$ on mode frequencies). In this case it is possible to discriminate between stars with well-separated $Y$, such as 0.250 and 0.400, but it is not possible to discriminate between stars with a difference in $Y$ of 0.040. In order to differentiate between stars whose $Y$ differ by 0.040 the size of the uncertainties on the mode frequencies must be less than $0.005\,\rm\mu Hz$. A simple scaling implies this would require more than 60\,yr of continuous high-quality observations. However, we must remember that this is only a rough estimate and the true size of the errors will also depend on factors such as the signal-to-noise ratio of the data and the lifetimes of the oscillations. Furthermore, these factors also mean that the size of the errors is not uniform across the range of modes considered. The above estimate for the length of time required to discriminate between stars with $Y$ that differ by 0.040 is, therefore, a worst case scenario.

Although dependent on the helium abundance, the location of the acoustic glitch cannot be used (with the uncertainties assumed here) to discriminate between populations of stars with different $Y$. However, the results do imply that, using the amplitude of the signal at $\nu_{\textrm{\scriptsize{max}}}$, we will be able to discriminate between models with $Y=0.250$ and $0.400$ using 1460\,d of $\nu_\textrm{\scriptsize{max}}>50\,\rm\mu Hz$. Such a comparison may be important for testing scenarios of high helium enrichment, such as the enrichment that may occur from the ejecta of massive asymptotic giant branch stars \citep[see][and references therein]{Gratton2012}. Examples of split populations within globular clusters that have very different $Y$ are becoming more frequent. One prominent example is the globular cluster $\omega$ Centauri, the most massive globular cluster in the Milky Way, which is believed to contain at least two distinct stellar populations, one of which is assumed to have the primordial helium abundance ($Y=0.25$), another population within the cluster is believed to have $Y=0.38$ \citep{Piotto2005} and there is even the possibility of another metal-rich component that may have $Y$ as high as 0.40 \citep{Lee2005, Sollima2005}. There is also evidence for similarly split populations in, for example, NGC 2808 \citep{DAntona2004, Piotto2007} and NGC 6441 \citep[][and references therein]{Caloi2007}. In fact, although not always with such widely separated $Y$ as in the above examples \citep[e.g][and references therein]{Milone2009} multiple stellar populations have been observed in numerous globular clusters. At present no seismic data are available for clusters with distinct, well-separated helium abundance populations. However, if, in the future, such data do become available use of seismic techniques to distinguish between red giants with different helium abundances would be particularly useful given that spectroscopic determinations of $Y$ are not possible due to their low effective temperatures.

One cluster that may be observed in the near future by \textit{Kepler's} K2 mission \citep{Chaplin2013} is the M4 globular cluster. The helium abundance of stars in this cluster appears to be enhanced by approximately 0.04 compared with the primordial helium content of the Universe \citep{Villanova2012}. Although this enhancement is small with respect to the differences in $Y$ we can reliably detect here it will be interesting to verify whether the differences in $Y$ estimated from the morphology of the colour-magnitude diagram and from the spectroscopic data of horizontal branch stars are at least compatible with the asteroseismic data.

Finally, we note that the $3\sigma$ difference required here is both reasonable and yet stringent. However, one need not restrict oneself to definitively stating that stars do or do not have the same initial helium abundance. Instead, it would be possible to extend the work done here in a statistically rigorous manner to determine the likelihood that two stars have the same initial helium abundance. Application of our methodology could, therefore, be instrumental in discriminating between RGB stars of populations with different $Y$ within globular clusters.

\appendix\section{Example fits using radial and non-radial modes}
\begin{figure*}
\centering
  \includegraphics[width=0.4\textwidth, clip]{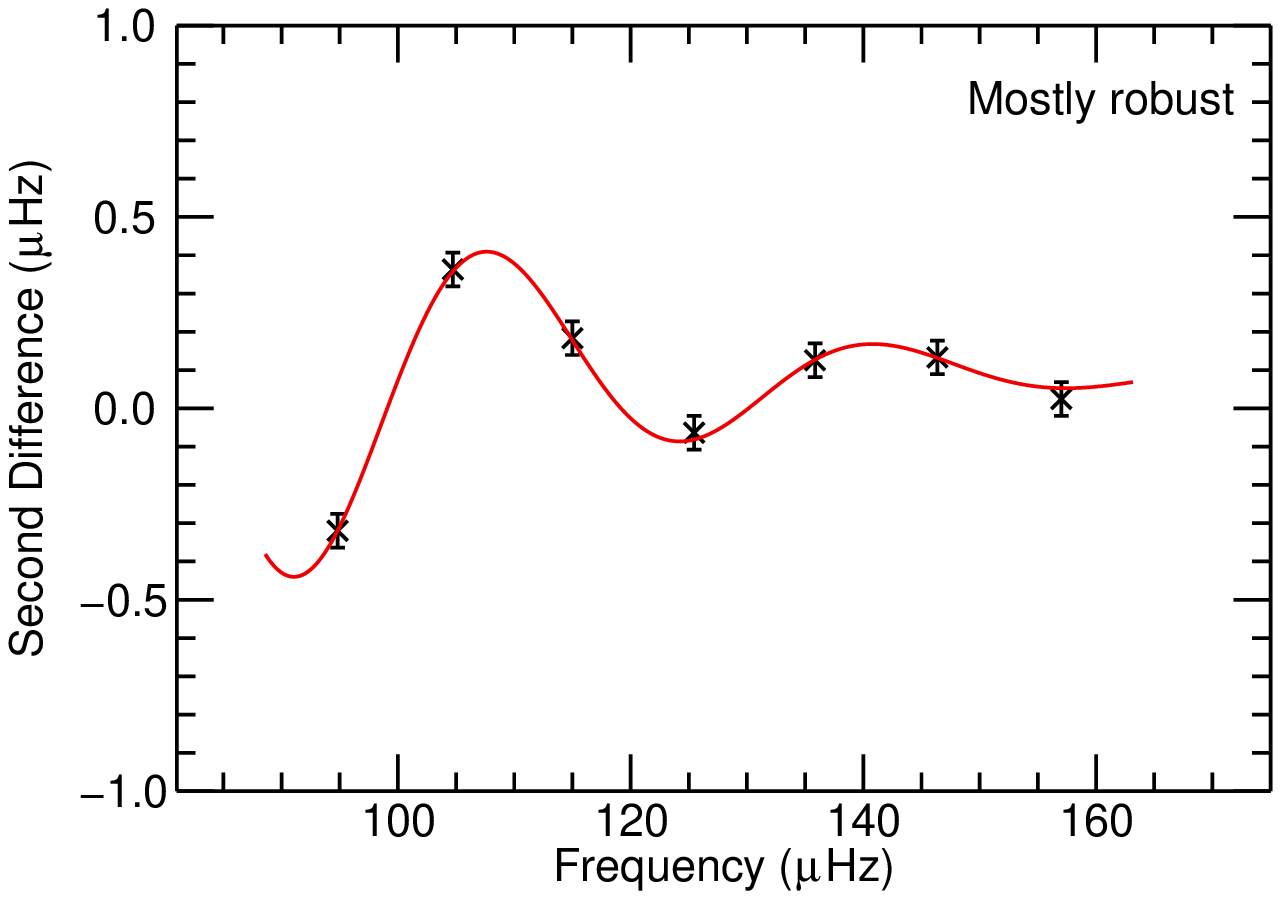}
  \includegraphics[width=0.4\textwidth, clip]{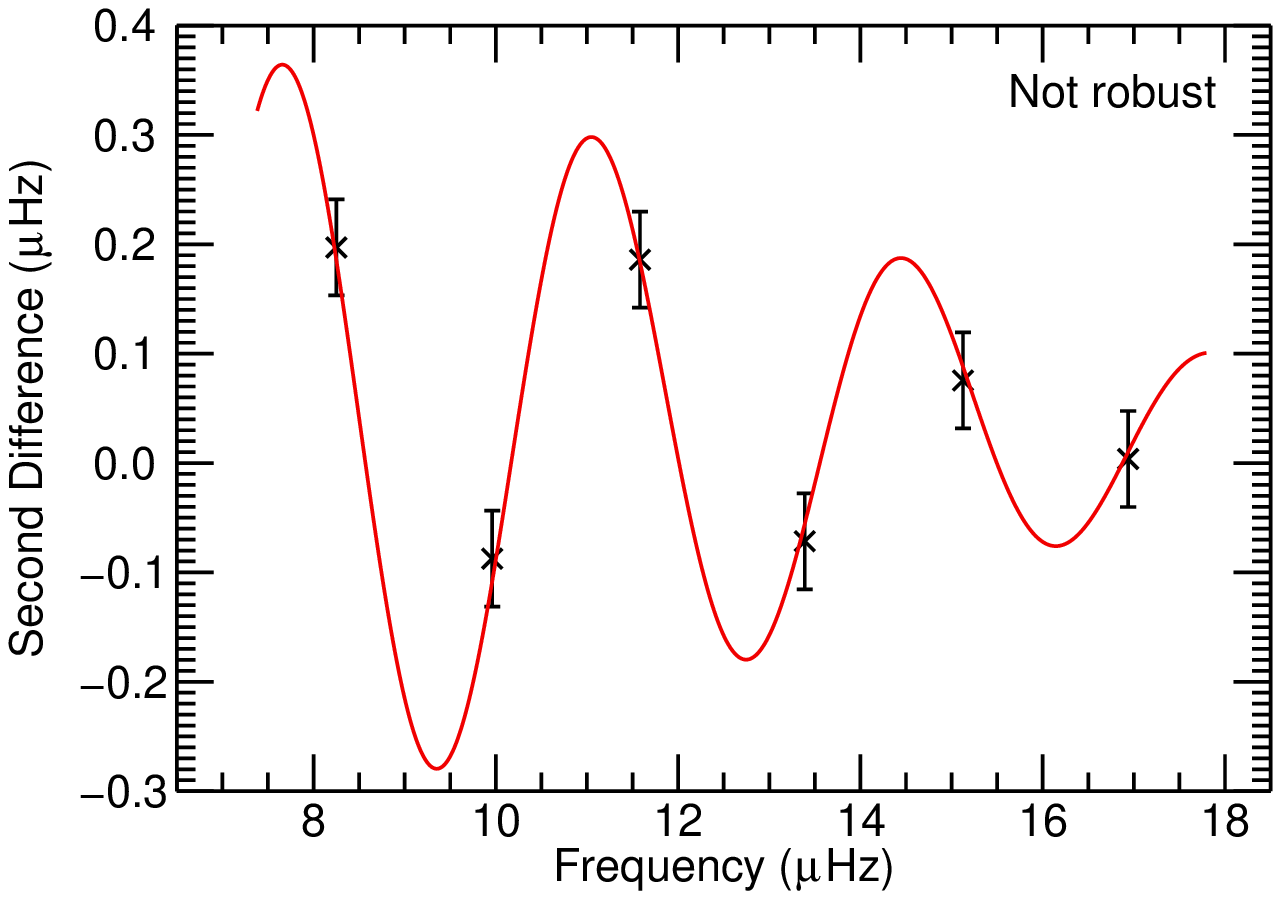}\\
  \includegraphics[width=0.4\textwidth, clip]{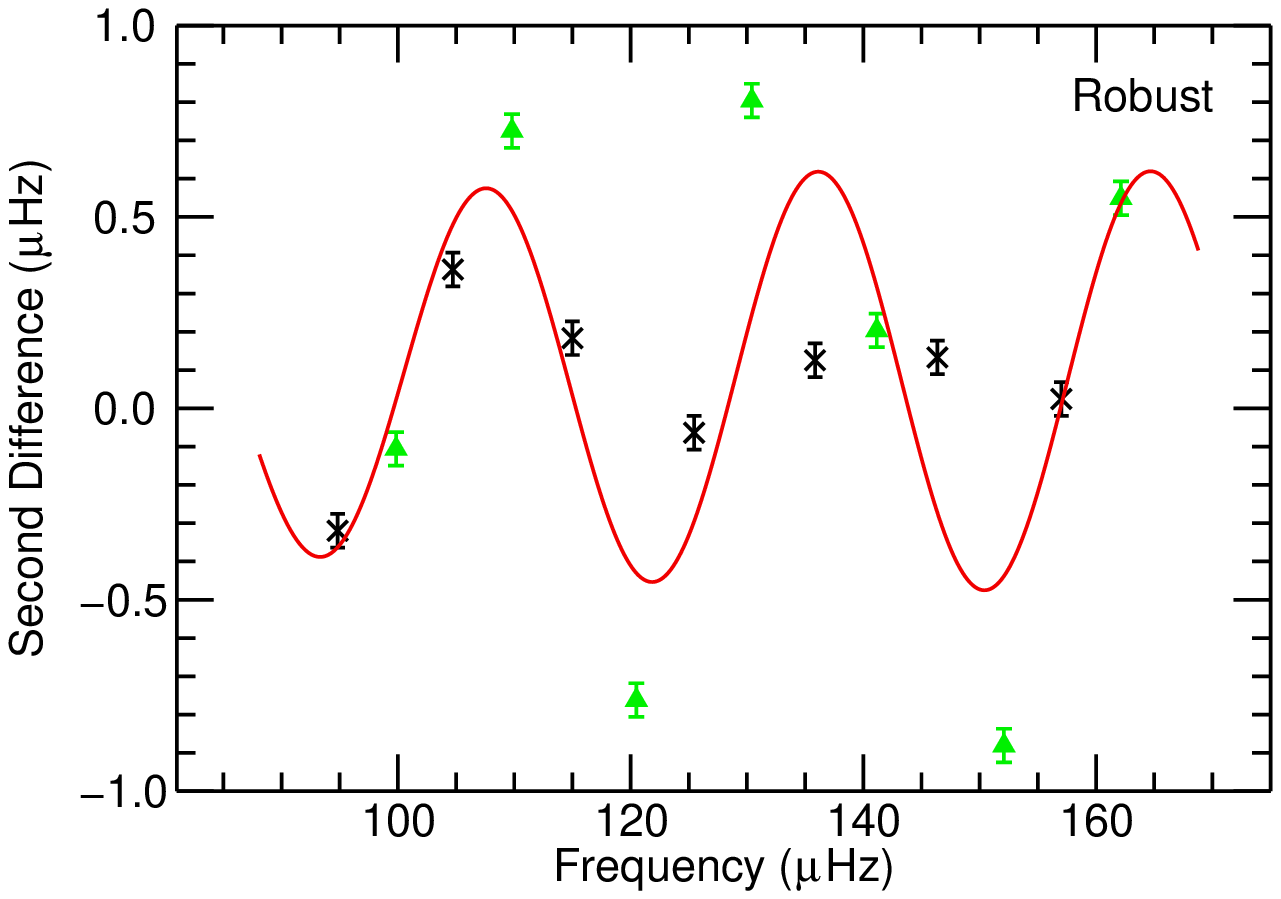}
  \includegraphics[width=0.4\textwidth, clip]{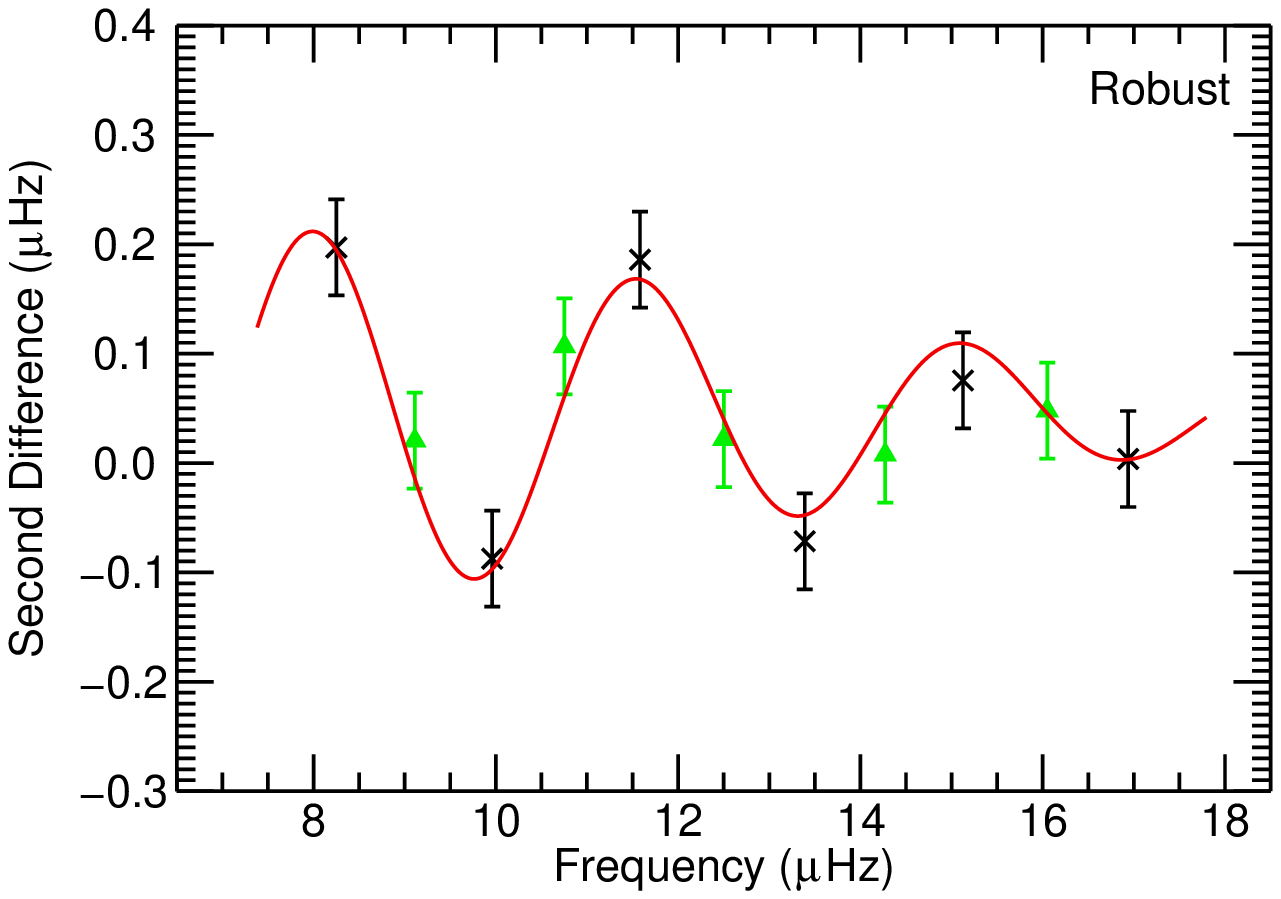}\\
  \includegraphics[width=0.4\textwidth, clip]{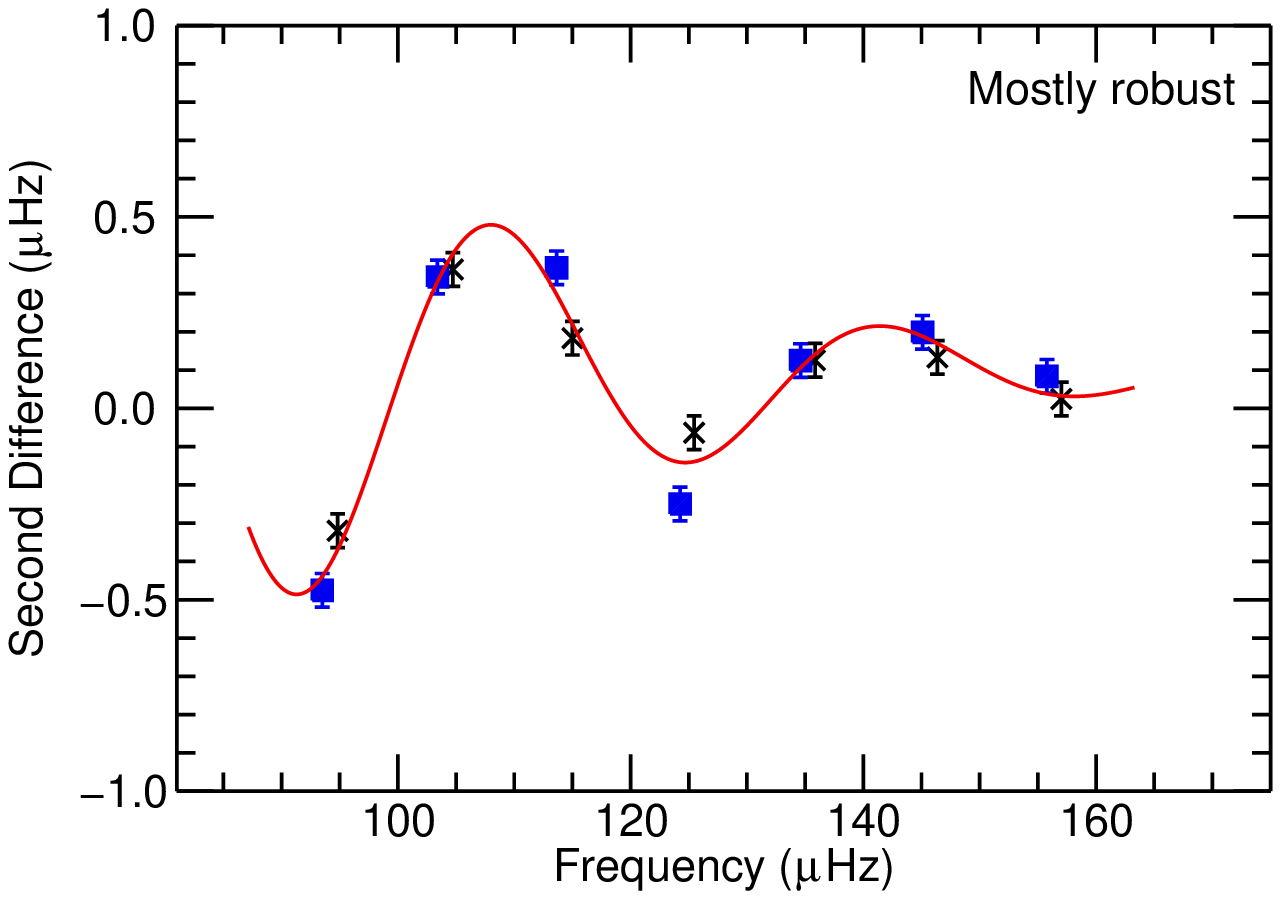}
  \includegraphics[width=0.4\textwidth, clip]{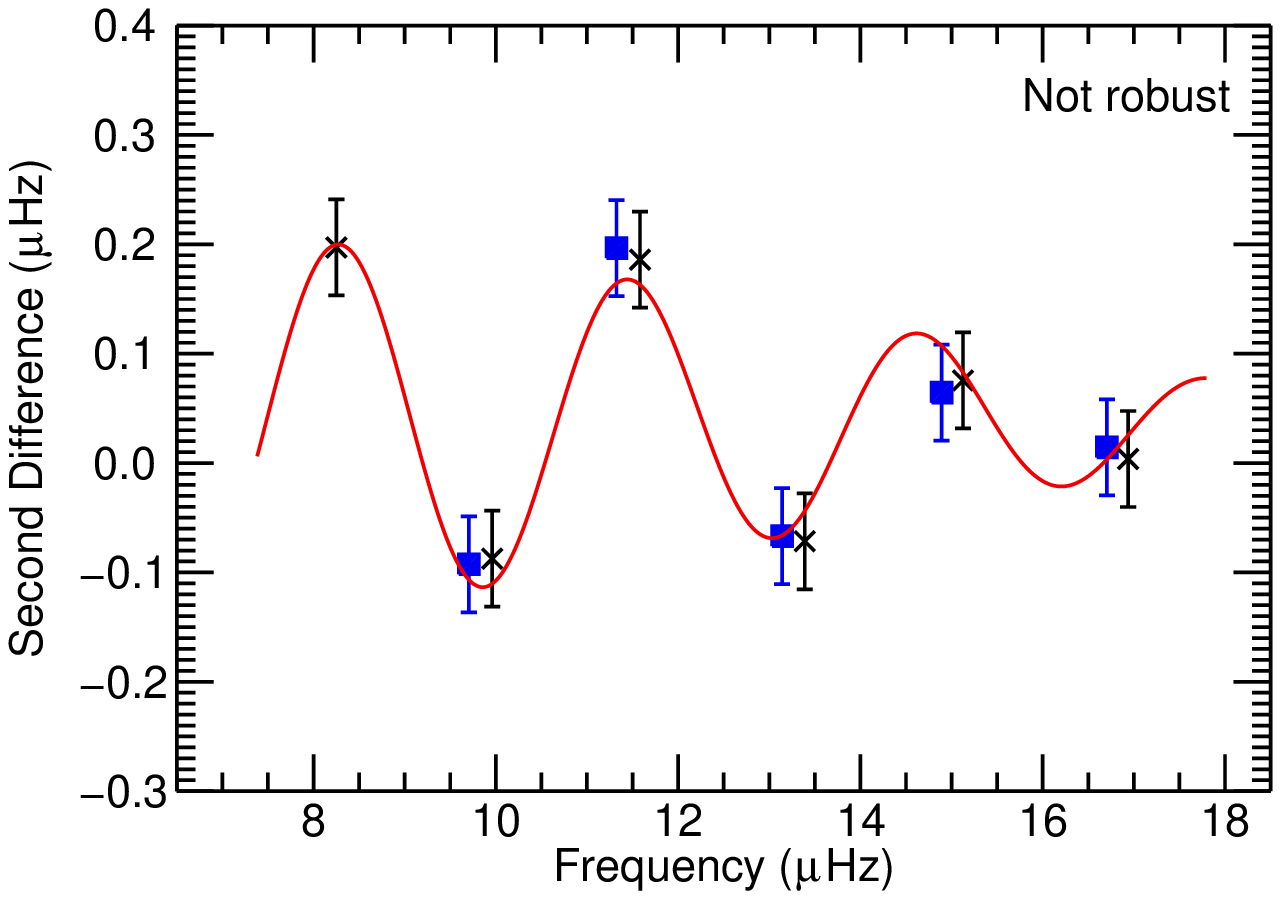}\\
  \includegraphics[width=0.4\textwidth, clip]{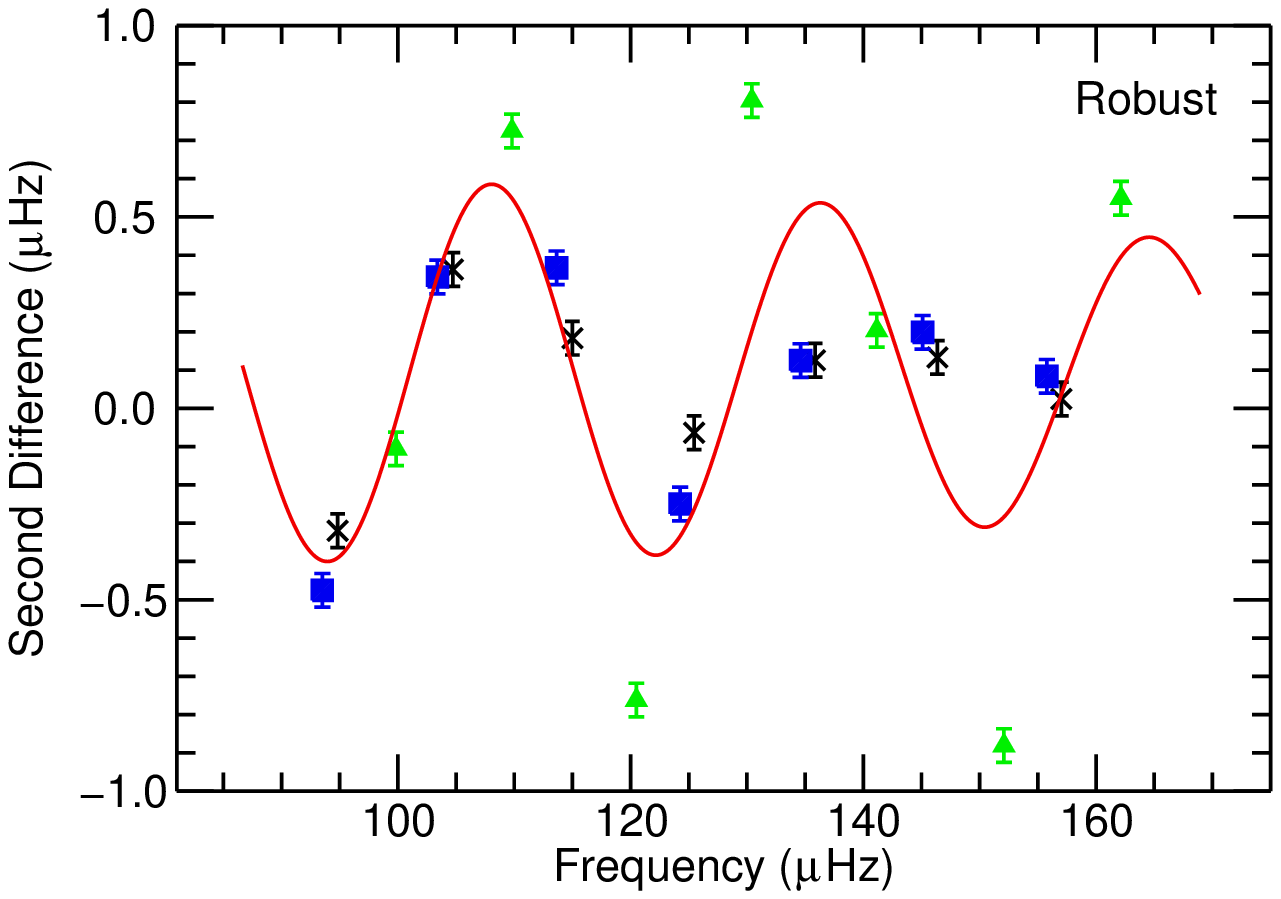}
  \includegraphics[width=0.4\textwidth, clip]{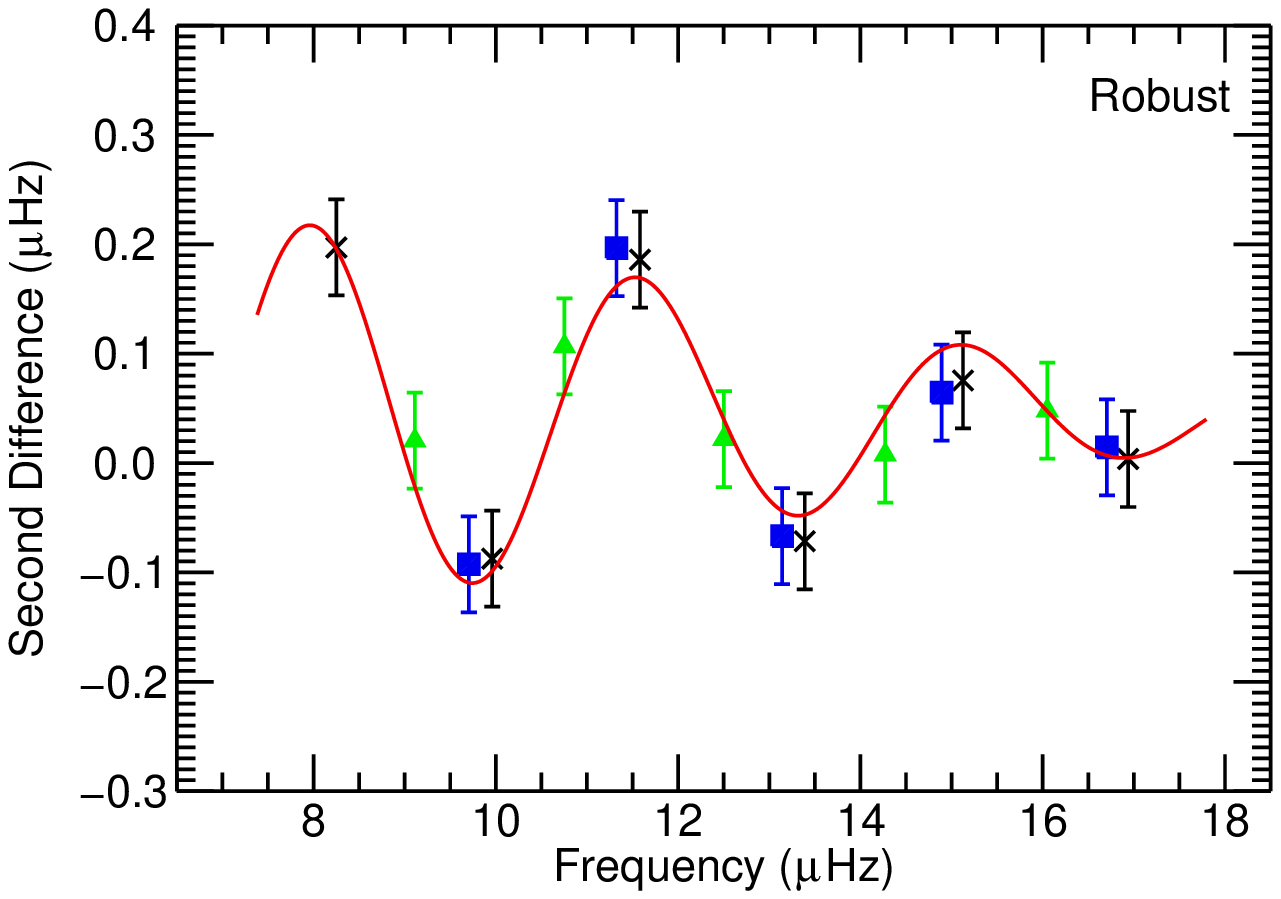}\\
  \caption{Example fits from two M1 model stars both with a mass of $1.5\,\rm M_\odot$, $Y=0.278$, and $Z=0.020$. Each panel in the left-hand column shows results for a star with $\nu_{\textrm{\scriptsize{max}}}=128.70\,\rm\mu Hz$, while each panel in the right-hand column shows results for a star with $\nu_{\textrm{\scriptsize{max}}}=12.71\,\rm\mu Hz$. First row: $l=0$ modes (black crosses) only used to fit the acoustic glitch. Second row: $l=0$ and $1$ modes (green triangles) used to fit the acoustic glitch. Third row: $l=0$ and $2$ modes (blue squares) used to fit the acoustic glitch. Fourth row: $l=0$, $1$, and $2$ modes used to fit the acoustic glitch.}\label{figure[eg fit l]}
\end{figure*}

Fig. \ref{figure[eg fit l]} shows examples of fits for two different stars when various combinations of $l$ were used. The robustness of the fits is indicated in the panels. We note that the fit for the model with $\nu_{\textrm{\scriptsize{max}}}=128.70\,\rm\mu Hz$ is robust when the $l=1$ modes are included despite the fact that the $l=1$ modes are mixed and therefore respond to the acoustic glitch in a different manner to the $l=0$ modes. This highlights the importance of visually checking each fit. We also note that in this example the fitted $t_{\textrm{\scriptsize{HeII}}}$ are  very different when the $l=1$ modes are included compared to the $t_{\textrm{\scriptsize{HeII}}}$ extracted when the $l=1$ modes are not included. This could possibly be used as a check as to whether it is appropriate to use non-radial overtones. The fits for the star with $\nu_{\textrm{\scriptsize{max}}}=12.71\,\rm\mu Hz$ were classified as not robust unless the $l=1$ modes were included, thus highlighting how the $l=1$ modes can be helpful for low $\nu_{\textrm{\scriptsize{max}}}$ stars.

\section*{Acknowledgements}
A-MB thanks the Institute of Advanced Study, University of Warwick for their support. WJC, YE, and AM acknowledge support from the UK Science and
Technology Facilities Council (STFC). Funding for the Stellar
Astrophysics Centre is provided by The Danish National Research
Foundation (Grant agreement no.: DNRF106).

\bibliographystyle{mn2e_new}
\bibliography{helium_abundance}

\end{document}